\documentclass[aps,prl,twocolumn,showpacs,superscriptaddress,superscriptaddress,superscriptaddress]{revtex4-1}

\usepackage[english]{babel}
\usepackage[pdftex]{graphicx}    % Option draft zeichnet für Bilder nur Rahmen mit Dateinamen
\usepackage{geometry}
\usepackage{graphicx}
\usepackage{hyperref}
\usepackage{floatrow}
\usepackage{subfigure}
\usepackage{floatflt}
\usepackage{amsmath}
\usepackage{floatflt,graphicx}
\usepackage{braket}
\usepackage{xcolor}
\usepackage{amssymb}
\usepackage{amsmath,amssymb,amsthm,mathrsfs,amsfonts,dsfont}
\usepackage{parskip}

\begin{document}

\title[]{Frequency Feedback for Two-Photon Interference from Separate Quantum~Dots}
%\title[]{Two-photon Interference from Separate, Frequency-Stabilized Quantum Dots}
%\title[]{Active Frequency Feedback for Two-Photon Interference from artificial Rb Atoms}
%Two-photon interference at Rb Transitions using frequency-locked Quantum Dots
%Active Frequency Feedback for two-photon interference between artificial Rb atoms
%Interference of single photons from two remote, frequency-locked quantum dots
%Active frequency feedback for indistinguishable photons from solid-state qubits

%\affiliation{Institut f{\"u}r Angewandte Physik, Universit{\"a}t Bonn, Wegelerstra{\ss}e 8, 53115~Bonn, Germany}
\author{Michael~\surname{Zopf}}
\affiliation{These authors contributed equally to this work.}
\affiliation{Institute for Integrative Nanosciences, Leibniz IFW Dresden, Helmholtzstra{\ss}e~20, 01069~Dresden, Germany}
\author{Tobias~\surname{Macha}}\email{macha@iap.uni-bonn.de}
\affiliation{These authors contributed equally to this work.}
\affiliation{Institut f{\"u}r Angewandte Physik, Universit{\"a}t Bonn, Wegelerstra{\ss}e 8, 53115~Bonn, Germany}
\author{Robert~\surname{Keil}}
\affiliation{Institute for Integrative Nanosciences, Leibniz IFW Dresden, Helmholtzstra{\ss}e~20, 01069~Dresden, Germany}
\author{Eduardo~\surname{Uru{\~n}uela}}
\affiliation{Institut f{\"u}r Angewandte Physik, Universit{\"a}t Bonn, Wegelerstra{\ss}e 8, 53115~Bonn, Germany}
\author{Yan~\surname{Chen}}
\affiliation{Institute for Integrative Nanosciences, Leibniz IFW Dresden, Helmholtzstra{\ss}e~20, 01069~Dresden, Germany}
\author{Wolfgang~\surname{Alt}}
\affiliation{Institut f{\"u}r Angewandte Physik, Universit{\"a}t Bonn, Wegelerstra{\ss}e 8, 53115~Bonn, Germany}
\author{Lothar~\surname{Ratschbacher}}
\affiliation{Institut f{\"u}r Angewandte Physik, Universit{\"a}t Bonn, Wegelerstra{\ss}e 8, 53115~Bonn, Germany}
\author{Fei~\surname{Ding}}\email{f.ding@fkp.uni-hannover.de}
\affiliation{Institute for Integrative Nanosciences, Leibniz IFW Dresden, Helmholtzstra{\ss}e~20, 01069~Dresden, Germany}
\affiliation{Institut f{\"u}r Festk{\"o}rperphysik, Leibniz Universit{\"a}t Hannover, Appelstra{\ss}e~2, 30167~Hannover, Germany}
\author{Dieter~\surname{Meschede}}
\affiliation{Institut f{\"u}r Angewandte Physik, Universit{\"a}t Bonn, Wegelerstra{\ss}e 8, 53115~Bonn, Germany}
\author{Oliver G.~\surname{Schmidt}}
\affiliation{Institute for Integrative Nanosciences, Leibniz IFW Dresden, Helmholtzstra{\ss}e~20, 01069~Dresden, Germany}
\affiliation{Material Systems for Nanoelectronics, Technische Universit{\"a}t Chemnitz, 09107~Chemnitz, Germany}
%Technische Universit{\"a}t 

\begin{abstract}
We employ active feedback to stabilize the frequency of single photons emitted by two separate quantum dots to an atomic standard. The transmission of a single, rubidium-based Faraday filter serves as the error signal for frequency stabilization to less than 1.5~\% of the emission linewidth. Long-term stability is demonstrated by Hong-Ou-Mandel interference between photons from the two quantum dots. The observed visibility of $V_{\mathrm{lock}}=(41 \pm 5)$~\% is limited only by internal dephasing of the dots. Our approach facilitates quantum networks with indistinguishable photons from distributed emitters.
\end{abstract}

\maketitle

Sources of indistinguishable photons play a fundamental role in concepts for photonic quantum computing~\cite{Obrien2007}, quantum communication~\cite{Gisin2007,Riedmatten2005} and distributed quantum networks~\cite{Pan2012}. Semiconductor quantum dots (QDs), also called `artificial atoms', are excellent candidates for the generation of single photons~\cite{Michler2000,Santori2001,He2013} and entangled photon pairs~\cite{Benson2000,Stevenson2006,Akopian2006,Hafenbrak2007,Muller2009,Dousse2010}. In recent years, major improvements of photon indistinguishability and source brightness have been accomplished~\cite{Somaschi2016,Ding2016}. In particular, GaAs/AlGaAs QDs enable pure single-photon and highly-entangled photon pair emission at rubidium D$_1$/D$_2$ transitions~\cite{Keil2017,Atkinson2012,Kumar2011}. Moreover, lifetime-limited emission from this type of QDs has been shown~\cite{Jahn2015}.

The emission frequency of a QD is sensitive to several external pertubations, including temperature~\cite{Giesz2015,Ding2016} as well as electric~\cite{Bennett2010,Ghali2012,Zhang2017}, magnetic~\cite{Stevenson2006,Pooley2014} and strain fields~\cite{Ding2010a,Zhang2015,Chen2016}. While these phenomena lead to spectral wandering of the QD emission over long timescales, they simultaneously provide means to fine-tune and match the emission frequencies using active frequency feedback~\cite{Metcalfe2009,Prechtel2013,Akopian2013}.

Here, we demonstrate a technique to counteract drifts by applying frequency feedback via strain tuning~\cite{Ding2010a,Trotta2012} of the host substrates of two separate QDs. Our frequency discrimination is achieved by measurements of the emitted single photon signal only, and with respect to an atomic rubidium standard. The QD emission is thereby matched to the Rb~D$_1$ transition at 795~nm, which is an attractive feature for efficient `quantum hybrid systems'~\cite{Rakher2013,Wolters2017}. The implementation of a common and reproducible standard paves the way towards quantum networks with distributed, indistinguishable solid-state emitters.

\begin{figure*}[ht]
	\includegraphics[width=1\textwidth]{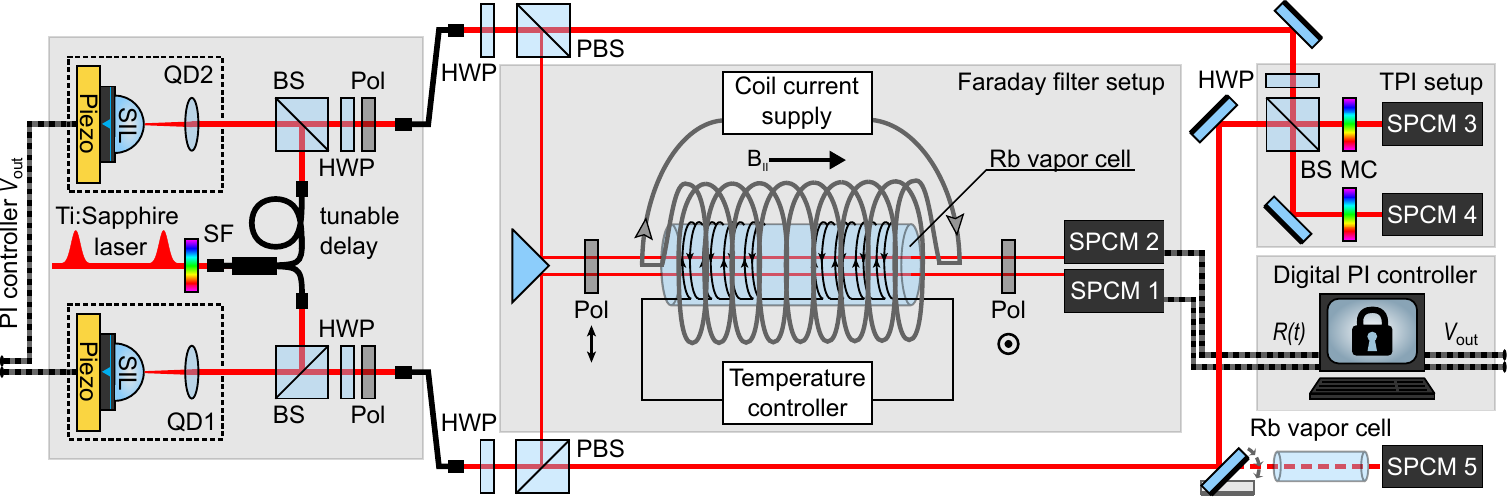}
	\caption[Setup]{Experimental setup for two-photon interference between separate, frequency-stabilized quantum dots (QDs). A pulsed Ti:sapphire laser resonantly excites the biexciton (XX) state of QD1 and QD2, positioned in separate cryostats at 4~K. Grating-based spectral filtering (SF) is applied to reduce the laser linewidth. A fiber-integrated, tunable delay adjusts the excitation to match arrival times of XX photons at the subsequent two-photon interference (TPI) setup. Both QDs are mounted on piezo-electric actuators for strain-induced emission frequency control. In each setup, the XX photons are collected using a confocal microscope, with the addition of a solid immersion lens (SIL) for enhanced extraction and a half-wave plate (HWP) and polarizer (Pol) for polarization filtering. A fraction of the signal is branched off by a HWP and a polarizing beam splitter (PBS) and sent through a Faraday filter setup. It consists of a heated Rb vapor cell in a longitudinal magnetic field, enclosed by two crossed polarizers. A coil current supply and a temperature controller enable tuning of the filter transmission features (see Fig.~\ref{fig:Faraday}). The transmitted XX photons are detected by single-photon counting modules (SPCMs) as signal inputs $R(t)$ for two digital PI controllers. Feedback voltages $V_\mathrm{out}$ are generated and applied to the piezoelectric actuators for QD frequency stabilization. An additional Rb vapour cell in the signal arm of QD1 permits characterization of frequency drifts independent of the Faraday filter. The remaining XX photon streams are sent to the TPI setup, consisting of a beam splitter (BS), monochromators (MCs) and SPCMs. A HWP in one input arm is used to set the photon (in)distinguishability with respect to the polarization state.} 
\label{fig:Setup}
\end{figure*}

In the following, we introduce the experimental setup and characterize the spectral quality of the QD emission, the frequency discriminator and the feedback technique. As a benchmark, we show an improved long-term two-photon interference (TPI) visibility of the frequency-stabilized QDs in a Hong-Ou-Mandel experiment~\cite{Hong1987,Patel2010,Flagg2010,Giesz2015}, in which only a small fraction of the emitted photon fluxes is required for locking.

The GaAs/AlGaAs QDs were grown by solid-source molecular beam epitaxy and \textit{in situ} Al droplet etching~\cite{Keil2017} and emit close to the rubidium D$_1$ transitions. Several QD-containing nanomembranes are obtained using wet chemical etching and are bonded to a piezoelectric actuator (0.3~mm PMN-PT) via a flip-chip transfer process~\cite{ZhangY2016}. Precise emission wavelength control is then achieved by applying a voltage to the actuator.

\begin{figure*}[ht]
	\includegraphics[width=1\textwidth]{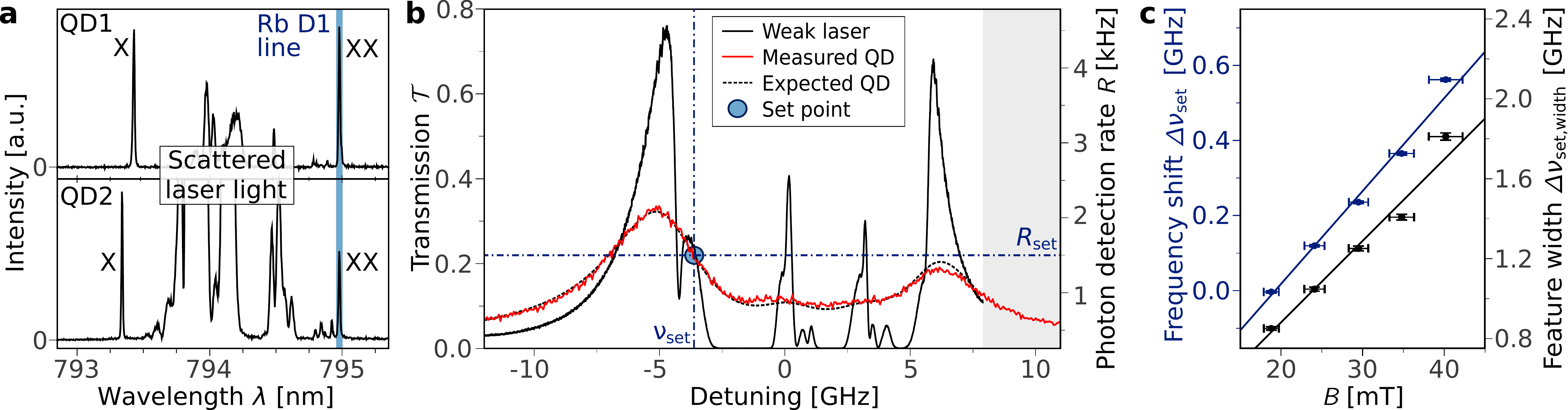}%
	\caption{(a) Emission spectra of two separate GaAs/AlGaAs QDs, whose emission frequencies are tuned in resonance with the Rb~D$_1$ transition using piezo-induced strain fields. A two-photon excitation scheme resonantly addresses the biexciton state, which decays via the exciton state (X) by emitting two consecutive photons (XX and X). The scattered laser is partially suppressed using notch filters. 
	(b) Faraday-filter transmission spectra measured with a weak, narrow-band laser (solid black line) and the frequency-tuned QD2 (red line). A convolution of the laser transmission with the spectral emission profile $f(\nu)$ of QD2 (dashed black line) is used to model the QD transmission. The detuning is given with respect to the weighted line center of the rubidium D$_1$ transitions. The set point for the TPI measurement (Fig.~\ref{fig:TPI}) with the reference photon rate $R_{\mathrm{set}} = R(\nu_\mathrm{set})$ is highlighted. Only relatively weak photon streams (here: $6000$ photons per second) are required to stabilize the frequency.
	(c) Tuning the magnetic field $B_\parallel$ allows to shift the filter's transmission peak, and thus the frequency set point of the stabilized emitter, to a desired frequency near the atomic resonance of the rubidium D$_1$ line with 24.6~MHz/mT. Simultaneously, the width of the transmission peak changes with 40.8~MHz/mT and can be adjusted to match the linewidth of the QD. The rubidium vapor cell was stabilized to a temperature of $85~^\circ\text{C}$.}
	\label{fig:Faraday}
\end{figure*}

The experimental setup is shown in Fig.~\ref{fig:Setup}. The QD samples (QD1 \& 2) are placed in two separate He cryostats at 4~K. A Ti:sapphire laser with $3$~ps pulse length and $76$~MHz repetition rate is sent to a grating-based pulse-shaping setup for spectral narrowing. The pulses are used to excite both QDs to the biexciton state (XX) in a resonant two-photon $\pi$-pulse condition~\cite{MullerM2014}. The XX photons at 795~nm are spectrally separated from laser and emission background using bandpass and notch filters. The fine-structure splitting of the exciton state (X) leads to two cross-polarized XX decay channels, of which only one is selected by polarization filtering. The respective emission spectra are shown in Fig.~\ref{fig:Faraday}a.

The lifetime of the XX state is determined for both QDs by a fluorescence decay measurement, revealing $T^{(\mathrm{QD1})}_{1}= (155 \pm 1)$~ps and $T^{(\mathrm{QD2})}_{1}= (187 \pm 1)$~ps. Using a Michelson interferometer, the coherence time $T_{2}$ and thus the Lorentzian linewidth $\Delta\nu$ of the emitted photons is determined. QD1 exhibits values of $T^{(\mathrm{QD1})}_{2}= (153 \pm 1)$~ps and $\Delta\nu^{(\mathrm{QD1})}= (2.08 \pm 0.01)$~GHz, and QD2 of $T^{(\mathrm{QD2})}_{2}= (123 \pm 4)$~ps and $\Delta\nu^{(\mathrm{QD2})}= (2.59 \pm 0.08)$~GHz. The calculated indistinguishability $I = T_2/2T_1$~\cite{Bylander2003} of the photons from each individual photon stream is therefore $I^{(\mathrm{QD1})}= (49.4 \pm 0.5)$~\% and $I^{(\mathrm{QD2})}= (32.9 \pm 1.1)$~\%.
 
Each QD emission is coupled into a single-mode fiber, delivering a photon rate of $30$~kcps. One part of each single-photon stream is sent to the TPI setup. It consists of a 50:50 non-polarizing beam splitter, followed by monochromators for further spectral filtering and single-photon counting modules (SPCMs) in each output arm.

As opposed to other feedback schemes~\cite{Metcalfe2009,Prechtel2013,Akopian2013}, a signal for frequency discrimination is provided by the Faraday effect. As depicted in Fig.~\ref{fig:Setup}, parts of the photon streams are directed to a \textit{Faraday anomalous dispersion optical filter} (FADOF) setup~\cite{Widmann2015}, which consists of a heated, natural-abundance rubidium vapor cell inside a solenoid, enclosed by crossed polarizers~\cite{Zielinska2011}. Therefore off-resonant background signals are efficiently suppressed, while on-resonance photons are transmitted and detected by SPCMs. The expected transmission ~$\mathcal{T}_{\mathrm{QD}}$ is given by a convolution of a narrow-band, weak laser transmission~$\mathcal{T}_{\mathrm{L}}$ with the spectral emission profile $f(\nu)$ of the QD: $\mathcal{T}_{\mathrm{QD}}(\nu)=(\mathcal{T}_{\mathrm{L}}*f)(\nu)$. Fig.~\ref{fig:Faraday}b shows the expected $\mathcal{T}_{\mathrm{QD2}}$ together with the measured frequency-tuned transmission of QD2. For a transmission peak close to the desired set frequency $\nu_{\mathrm{set}}$, the slope around $\nu_{\mathrm{set}}$ serves as the error signal for frequency stabilization. Changes in frequency are directly translated to a variation of the FADOF transmission. The latter also depends on both temperature $T$ and axially-applied magnetic field $B_\parallel$~\footnote{The software \textit{ElecSus}~\cite{Zentile2015} is used to calibrate the conversion from coil current to magnetic field.}, which provides a possibility to shift the transmission peak to a desired frequency near the atomic resonance of the rubidium D$_1$ line ($5^2S_{1/2}\rightarrow 5^2P_{1/2}$, 795 nm), as demonstrated in Fig.~\ref{fig:Faraday}c. Simultaneously, the width of the transmission peak can be adjusted to match the linewidth of the QD.
%The rubidium vapor cell was stabilized to a temperature of $85~^\circ\text{C}$.
%With increasing magnetic field, a given transmission shifts with 24.6~MHz/mT, as demonstrated in Fig.~\ref{fig:Faraday}c. This corresponds to an effective change of the set point frequency $\nu_\mathrm{set}$. Simultaneously, the width of the transmission peak $\Delta\nu_\mathrm{set,width}$ can be adjusted to match the linewidth of the QD with 40.8~MHz/mT. The rubidium vapor cell was stabilized to a temperature of $85~^\circ\text{C}$.

The SPCM photon detection rate $R_{\mathrm{set}}=R(\nu_{\mathrm{set}})$ serves as the reference for the frequency feedback. The rate $R$ of photon events at the SPCM can be written as
\begin{equation}
%\begin{aligned}
R(t) = R(\nu(t))\equiv \mathcal{T}_\mathrm{QD}(\nu(t)) \cdot R_\mathrm{QD} , 
%\end{aligned}
\label{eq:rate}
\end{equation}
which depends on the time-varying center frequency $\nu(t)$ of the QD's spectral emission profile. Here it is assumed that $R_{\mathrm{QD}}$ already describes the QD emission rate reduced by the finite quantum efficiency of the detector. By inverting Eq.~\ref{eq:rate}, the instantaneous frequency deviation from the set point $\Delta\nu(t)\equiv\nu(t)-\nu_\mathrm{set}$ can be determined, using the observed detection rate $R(t)$. In practice, deviations from the set point are kept small by the feedback loop and the linearized relation
\begin{equation}
\Delta\nu \approx \frac{1}{\left. \frac{dR}{d\nu}\right|_{\nu_\mathrm{set}}} \Delta R
\label{eq:slope}
\end{equation} 
with $\Delta R = R(t) - R_\mathrm{set}$ provides a good approximation.\\
In order to obtain an error signal for feedback, a simple, empirical algorithm is implemented to estimate the underlying scattering rate at any point in time. Motivated by the fact that photon events that lie further in the past convey less information and should thus be given lower weight with time, an exponential smoothing filter is chosen to estimate the count rate $R(t)$.
The digital implementation is similar to a first order low pass filter and described by the pseudo-code
\begin{equation}
\begin{aligned}
R_{\mathrm{estimate},n+1}&=R_{\mathrm{estimate},n} \cdot d  +  \mathcal{B} \cdot i\\
\\
\text{where \  } \mathcal{B}&=\left\{\begin{array}{cl} 1, & \mbox{if a photon arrived }\\ 0, & \mbox{else,} \end{array}\right. 
\end{aligned}
\label{eq:algorithm}
\end{equation}
with the decrement $d = e^{-\tau_\mathrm{cycle}/\tau_\mathrm{filter}}$ and the increment $i = (1-d)/\tau_\mathrm{cycle}$. Here, $\tau_\mathrm{cycle}$ and $\tau_\mathrm{filter}$ denote the cycle time of the digital loop and the chosen integration time of the filter, respectively. Instead of using a discrete averaging window~\cite{Metcalfe2009}, our algorithm represents an infinite impulse response filter and thus features a smooth frequency response.

There are two important aspects for rate-based frequency estimations: The first one is the correct detection of variations in the scattering rate $R(t)$ from the stochastic train of photon detection events observed by the SPCM. We measure the free-running QD frequency-noise power spectral densities~\cite{Kuhlmann2013} on the rate $R(t)$ to determine the frequency at which the QD $1/f$-noise is exceeded by detection shot noise. Then the feedback bandwidth of the control system is set to a frequency well below.\\
The second aspect is the distinction between rate variations due to frequency drifts and due to intensity changes in the QD emission. The latter can in principle be compensated by adjusting the rate $R_{\mathrm{set}}$ with respect to a rate measurement before the Faraday filter. However, in our experiment the excitation laser background is varying strongly over time and prevents intensity stabilization. The influence on the rate after the filter, i.e. with additional frequency filtering, is small in comparison. QD intensity fluctuations due to sample drifts are taken into account by selecting data windows in which the count rate after the TPI setup is stable.

The rate estimation algorithm as well as a subsequent standard digital proportional-integral (PI) controller are implemented on a \textit{field programmable gate array} (FPGA) \footnote{National Instruments NI PXI-7842R card}
%~\footnote{The card features a maximum (logic) clock rate of $40\,$MHz and a maximum update rate of $1\,$MHz for the 16bit DAC (Digital-to-Analog Converter) with $-10\,$V to $+10\!$~V output range} 
using LabView. A graphical user interface allows for monitoring and controlling the frequency feedback loop on a standard Windows PC. The generated correction signal is sent to the strain-tuning piezoelectric actuator beneath the QD via a high-voltage amplifier.\\
%A NI SBC-68 connection box receives the TTL signals from the SPCM and sends the correction signal via a high-voltage amplifier to the strain-tuning piezo beneath the quantum dot. In \textit{scan mode}, a linear frequency scan is applied to obtain the FADOF transmission of the quantum dot photons in Fig.~\ref{fig:Faraday}a, in \textit{lock mode} the reference frequency rate $R(\nu_{\mathrm{set}})$ is kept, see Fig.~\ref{fig:TPI}c.

\begin{figure*}[ht]
	\includegraphics[width=1\textwidth]{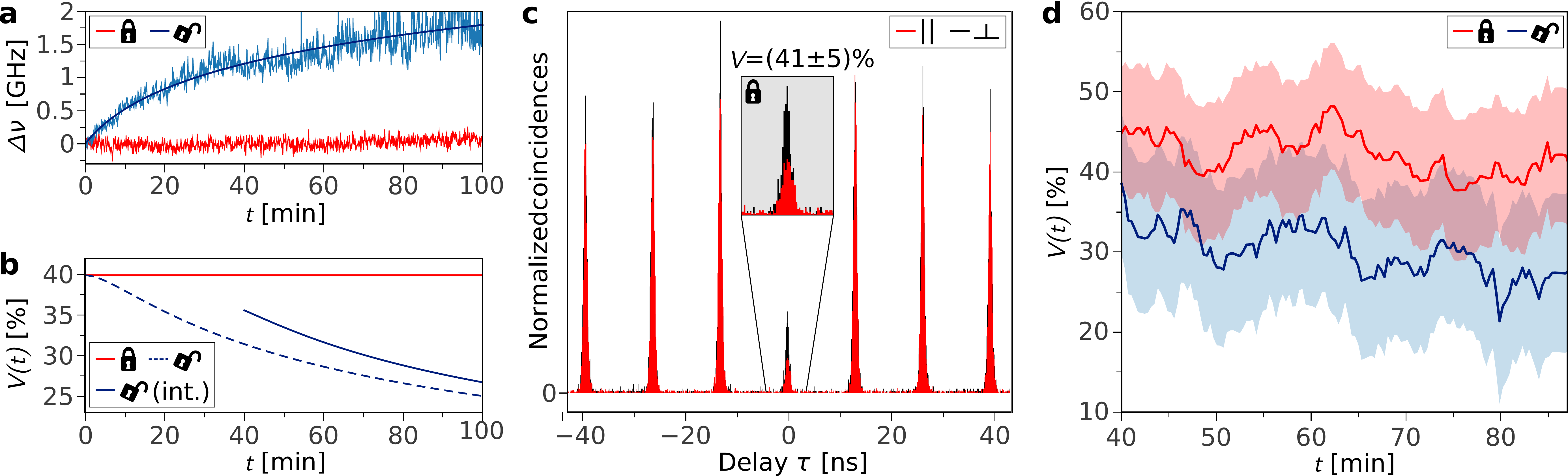}
	\caption{(a) Relative emission-frequency drift over time for QD1 in frequency-locked and free-running state, obtained by an out-of-loop measurement of the transmission through a separate Rb vapor cell. In the stabilized state the frequency is kept constant within a deviation of less than $30$~MHz, which is below 1.5~\% of the emission linewidth. The free-running state reveals a frequency drift due to piezo creep, which is fitted by a logarithmic function.
	(b) Theoretically expected evolution of the TPI visibility $V(t)$ considering the frequency drift shown in (a). As parameters serve the experimentally determined lifetimes $T_1$ and coherence times $T_2$ of the two separate QDs. While the visibility stays constant in the frequency-locked case (red line), for free-running QDs it drops from $V_{\mathrm{free}}(t=0~\mathrm{min})=$~40~\% to $V_{\mathrm{free}}(t=100~\mathrm{min})=$~25~\% (dashed line). The solid blue line represents a coincidence integration window of 40~min as used for the experiment in (d).
	(c) Two-photon interference measurement between two frequency-stabilized, separate QDs, showing the normalized coincidences versus the delay time $\tau$. The black curve corresponds to perpendicular polarizations of the photons impinging on the beam splitter. The red curve for parallel polarizations depicts a clear reduction of coincidences at $\tau\approx 0$. An interference visibility of $V_{\mathrm{lock}}=(41 \pm 5)$~\% is obtained. A similar measurement of the visibility with free-running QDs (not shown) results in $V_{\mathrm{free}}=(31 \pm 7)$~\%.
		(d) Measurement of the interference visibility over time for both free-running and frequency-locked QDs. Each data point corresponds to the coincidences obtained for the previous 40~min. The shaded areas are the respective uncertainties based on Poisson counting statistics. At any measurement time the visibility is higher for frequency-locked QDs than for free-running QDs.}
	\label{fig:TPI}
\end{figure*}

In this work, the strain-tuning mechanism is on the one hand used to obtain emission frequency control, on the other hand it is also the most prominent disturbance for frequency stability itself. Due to piezo creep, a certain set voltage on the piezoelectric actuator will not result in a constant strain in the QD membrane. The strain will slightly change over time and therefore result in a frequency drift, which is compensated by the implemented stabilization.
For locking the QD emission frequencies, count rates of $R_{\mathrm{set,QD1}}=3600$~cps and $R_{\mathrm{set,QD2}}=1500$~cps are used, which are significantly smaller than typical emission rates. At the set point of $R_{\mathrm{set,QD2}}$, depicted in Fig.~\ref{fig:Faraday}b, the feedback bandwidth is limited to around 30~mHz by adjusting the PI parameters.
%~\footnote{The GUI also allows to record Bode plots, in which we have seen a cable capacitance-limited bandwidth of 4~kHz for higher rates.}.
Fig.~\ref{fig:TPI}a shows the frequency drift of QD1 for the frequency-locked and free-running case. It is determined by measuring the transmission of the photon stream through a separate, heated rubidium vapor cell, which constitutes an out-of-loop measurement of the frequency drift. Frequency-stabilization leads to a constant frequency within a deviation of $<30$~MHz~\footnote{Calculated using $\sqrt{\sigma_N^2-\overline{N}}$ to exclude the detection shot noise. $\overline{N}$ is the average count number for $\geq$0.5~s binning times and $\sigma_N$ is the corresponding standard deviation.}, which is less than 1.5~\% of the linewidths of the QDs ($\geq$~2~GHz). In the free-running case, the frequency detuning $\Delta\nu(t)$ increases over time, following a logarithmic law known for the displacement change due to piezo creep~\cite{Jung2000}:
\begin{equation}
\Delta\nu(t) = \Delta\nu_0 \cdot \left[1+\alpha \cdot \mathrm{log}_{10}(t-t_0)\right]
\end{equation}
Here, $\Delta\nu_0$ denotes the frequency detuning 1~minute after a certain voltage is applied to the piezo at a time $t_0$, and $\alpha$ describes the rate of the piezo creep, which depends on the applied voltage and the piezo load. The displayed data in Fig.~\ref{fig:TPI}a is in good agreement with the model. The resulting values for frequency drift over time are then used to calculate the theoretical TPI interference visibility for the locked and free-running QDs, taking the experimental parameters of the QD photons into account. Fig.~\ref{fig:TPI}b shows the expected visibility over time, assuming a frequency drift between the two QD emission frequencies as observed in Fig.~\ref{fig:TPI}a. The visibility is calculated by~\cite{Giesz2015}
\begin{equation}
V = \frac{\gamma_1\gamma_2}{\gamma_1+\gamma_2} \cdot \frac{\gamma_1+\gamma_2+\gamma^*_1+\gamma^*_2}{(2\pi\Delta\nu)^2 + (\gamma_1+\gamma_2+\gamma^*_1+\gamma^*_2)^2/4}
\end{equation}
with $\gamma_i = 1/T^{(\mathrm{QD}i)}_{1}$ denoting the radiative decay rate and $\gamma^*_i = (2/T^{(\mathrm{QD}i)}_{2} - \gamma_i)$ the pure dephasing rate for the different QDs ($i=$1,2). Perfect frequency stability results in a constant maximum visibility of $V=40$~\%, while for the measured piezo creep the theoretically expected visibility drops to $V=25$~\% at $t=$100~min.

In order to experimentally verify an improved long-term visibility under frequency stabilization, we compare the TPI of photons from two separate QDs in the frequency-locked and free-running state. For stabilized QDs, Fig.~\ref{fig:TPI}c shows the normalized coincidences of photons in the two beam splitter output ports versus the delay time $\tau$ between the recorded events. The polarization state between the interfering photons is controlled by a half-wave plate, which allows to set the polarizations of photons impinging on the beam splitter perpendicular or parallel to each other. The interference visibility $V$ is calculated by evaluating the peak areas $A_{\parallel}$ for parallel and $A_{\perp}$ for perpendicular polarizations at $\tau=0$:
\begin{equation}
V = \frac{A_{\perp} - A_{\parallel}}{A_{\perp}}
\end{equation}

A clear Hong-Ou-Mandel dip is observed, yielding an interference visibility of $V_{\mathrm{lock}}=(41 \pm 5)$~\% after dark count correction of the SPCMs ($R_\mathrm{dc,SPCM3} =$~104~cps, $R_\mathrm{dc,SPCM4} =$~134~cps). The visibility agrees well with the expected value of $V=40$~\% in Fig.~\ref{fig:TPI}b. Afterwards, a measurement with free-running QDs is performed. The visibility in that case decreases to $V_{\mathrm{free}}=(31 \pm 7)$~\%, due to piezo creep and other frequency perturbations. For ideal quantum emitters the ratio of the peak at $\tau=0$ compared to the neighboring peaks equals 0.5 for perpendicular photon polarizations~\cite{Santori2001}. Here, a lower ratio is observed, which can be attributed to blinking in the QD emission~\cite{Jons2017}. This effect happens at sub-microsecond timescales, to which the frequency stabilization is insensitive.
To further compare the two cases of TPI with and without frequency feedback, the interference visibility is measured as a function of time, as shown in Fig.~\ref{fig:TPI}d. Each respective data point corresponds to the coincidences obtained within the previous 40~min. Hence, the integration window is gradually shifted through the total measurement time of 87~min. The shaded areas display the respective uncertainties due to Poisson counting statistics. In the locked and free-running case, both QDs were frequency-matched at $t=$0~min. In the free-running case, frequency changes in the QD emission within the first integration window reduce the visibility already for the first data points (see also Fig.~\ref{fig:TPI}b).

In conclusion, we have verified that active frequency feedback is an attractive solution to maintain long-term indistinguishability of photons from separate solid-state emitters. Stable two-photon interference from separate quantum dots is achieved by strain-mediated frequency stabilization. Frequency fluctuations are suppressed to a negligible fraction of the emission linewidth. The rubidium-based Faraday filter offers a common frequency standard for distant nodes in a quantum network. Furthermore, matching the rubidium transitions is desirable for rubidium-based quantum memories as potential elements in quantum repeaters. Low filter losses and an efficient rate-estimation algorithm ensure frequency stabilization while using only a small fraction of the photon flux.\\
The presented experiment can be extended to stabilize the emission of entangled photons for realizing a stable Bell state measurement in entanglement swapping schemes. In the case of QDs this means maintaining a low fine structure splitting by separating emitted photons according to their polarization and using two orthogonal degrees of freedom for feedback, as available in anisotropic strain-tuning platforms~\cite{Chen2016}.

We acknowledge funding by the BMBF (Q.Com), the European Commission (HANAS) and the ERC (QD-NOMS). F.D. acknowledges the support of IFW Excellence Program. T.M. \& E.U. thank the BCGS for support. We thank the IAP workshop and the IFW cleanroom staff for technical support.


\begin{thebibliography}{48}%
\makeatletter
\providecommand \@ifxundefined [1]{%
 \@ifx{#1\undefined}
}%
\providecommand \@ifnum [1]{%
 \ifnum #1\expandafter \@firstoftwo
 \else \expandafter \@secondoftwo
 \fi
}%
\providecommand \@ifx [1]{%
 \ifx #1\expandafter \@firstoftwo
 \else \expandafter \@secondoftwo
 \fi
}%
\providecommand \natexlab [1]{#1}%
\providecommand \enquote  [1]{``#1''}%
\providecommand \bibnamefont  [1]{#1}%
\providecommand \bibfnamefont [1]{#1}%
\providecommand \citenamefont [1]{#1}%
\providecommand \href@noop [0]{\@secondoftwo}%
\providecommand \href [0]{\begingroup \@sanitize@url \@href}%
\providecommand \@href[1]{\@@startlink{#1}\@@href}%
\providecommand \@@href[1]{\endgroup#1\@@endlink}%
\providecommand \@sanitize@url [0]{\catcode `\\12\catcode `\$12\catcode
  `\&12\catcode `\#12\catcode `\^12\catcode `\_12\catcode `\%12\relax}%
\providecommand \@@startlink[1]{}%
\providecommand \@@endlink[0]{}%
\providecommand \url  [0]{\begingroup\@sanitize@url \@url }%
\providecommand \@url [1]{\endgroup\@href {#1}{\urlprefix }}%
\providecommand \urlprefix  [0]{URL }%
\providecommand \Eprint [0]{\href }%
\providecommand \doibase [0]{http://dx.doi.org/}%
\providecommand \selectlanguage [0]{\@gobble}%
\providecommand \bibinfo  [0]{\@secondoftwo}%
\providecommand \bibfield  [0]{\@secondoftwo}%
\providecommand \translation [1]{[#1]}%
\providecommand \BibitemOpen [0]{}%
\providecommand \bibitemStop [0]{}%
\providecommand \bibitemNoStop [0]{.\EOS\space}%
\providecommand \EOS [0]{\spacefactor3000\relax}%
\providecommand \BibitemShut  [1]{\csname bibitem#1\endcsname}%
\let\auto@bib@innerbib\@empty
%</preamble>
\bibitem [{\citenamefont {O'Brien}(2007)}]{Obrien2007}%
  \BibitemOpen
  \bibfield  {author} {\bibinfo {author} {\bibfnamefont {J.~L.}\ \bibnamefont
  {O'Brien}},\ }\href
  {http://science.sciencemag.org/content/318/5856/1567.abstract} {\bibfield
  {journal} {\bibinfo  {journal} {Science}\ }\textbf {\bibinfo {volume}
  {318}},\ \bibinfo {pages} {1567} (\bibinfo {year} {2007})}\BibitemShut
  {NoStop}%
\bibitem [{\citenamefont {Gisin}\ and\ \citenamefont {Thew}(2007)}]{Gisin2007}%
  \BibitemOpen
  \bibfield  {author} {\bibinfo {author} {\bibfnamefont {N.}~\bibnamefont
  {Gisin}}\ and\ \bibinfo {author} {\bibfnamefont {R.}~\bibnamefont {Thew}},\
  }\href {http://dx.doi.org/10.1038/nphoton.2007.22} {\bibfield  {journal}
  {\bibinfo  {journal} {Nat. Photon.}\ }\textbf {\bibinfo {volume} {1}},\
  \bibinfo {pages} {165} (\bibinfo {year} {2007})}\BibitemShut {NoStop}%
\bibitem [{\citenamefont {de~Riedmatten}\ \emph {et~al.}(2005)\citenamefont
  {de~Riedmatten}, \citenamefont {Marcikic}, \citenamefont {van Houwelingen},
  \citenamefont {Tittel}, \citenamefont {Zbinden},\ and\ \citenamefont
  {Gisin}}]{Riedmatten2005}%
  \BibitemOpen
  \bibfield  {author} {\bibinfo {author} {\bibfnamefont {H.}~\bibnamefont
  {de~Riedmatten}}, \bibinfo {author} {\bibfnamefont {I.}~\bibnamefont
  {Marcikic}}, \bibinfo {author} {\bibfnamefont {J.~A.~W.}\ \bibnamefont {van
  Houwelingen}}, \bibinfo {author} {\bibfnamefont {W.}~\bibnamefont {Tittel}},
  \bibinfo {author} {\bibfnamefont {H.}~\bibnamefont {Zbinden}}, \ and\
  \bibinfo {author} {\bibfnamefont {N.}~\bibnamefont {Gisin}},\ }\href
  {https://doi.org/10.1103/PhysRevA.71.050302} {\bibfield  {journal} {\bibinfo
  {journal} {Phys. Rev. A}\ }\textbf {\bibinfo {volume} {71}} (\bibinfo {year}
  {2005})}\BibitemShut {NoStop}%
\bibitem [{\citenamefont {Pan}\ \emph {et~al.}(2012)\citenamefont {Pan},
  \citenamefont {Chen}, \citenamefont {Lu}, \citenamefont {Weinfurter},
  \citenamefont {Zeilinger},\ and\ \citenamefont {Zukowski}}]{Pan2012}%
  \BibitemOpen
  \bibfield  {author} {\bibinfo {author} {\bibfnamefont {J.~W.}\ \bibnamefont
  {Pan}}, \bibinfo {author} {\bibfnamefont {Z.~B.}\ \bibnamefont {Chen}},
  \bibinfo {author} {\bibfnamefont {C.~Y.}\ \bibnamefont {Lu}}, \bibinfo
  {author} {\bibfnamefont {H.}~\bibnamefont {Weinfurter}}, \bibinfo {author}
  {\bibfnamefont {A.}~\bibnamefont {Zeilinger}}, \ and\ \bibinfo {author}
  {\bibfnamefont {M.}~\bibnamefont {Zukowski}},\ }\href
  {https://doi.org/10.1103/RevModPhys.84.777} {\bibfield  {journal} {\bibinfo
  {journal} {Rev. Mod. Phys.}\ }\textbf {\bibinfo {volume} {84}},\ \bibinfo
  {pages} {777} (\bibinfo {year} {2012})}\BibitemShut {NoStop}%
\bibitem [{\citenamefont {Michler}\ \emph {et~al.}(2000)\citenamefont
  {Michler}, \citenamefont {Kiraz}, \citenamefont {Becher}, \citenamefont
  {Schoenfeld}, \citenamefont {Petroff}, \citenamefont {Zhang}, \citenamefont
  {Hu},\ and\ \citenamefont {Imamoglu}}]{Michler2000}%
  \BibitemOpen
  \bibfield  {author} {\bibinfo {author} {\bibfnamefont {P.}~\bibnamefont
  {Michler}}, \bibinfo {author} {\bibfnamefont {A.}~\bibnamefont {Kiraz}},
  \bibinfo {author} {\bibfnamefont {C.}~\bibnamefont {Becher}}, \bibinfo
  {author} {\bibfnamefont {W.~V.}\ \bibnamefont {Schoenfeld}}, \bibinfo
  {author} {\bibfnamefont {P.~M.}\ \bibnamefont {Petroff}}, \bibinfo {author}
  {\bibfnamefont {L.}~\bibnamefont {Zhang}}, \bibinfo {author} {\bibfnamefont
  {E.}~\bibnamefont {Hu}}, \ and\ \bibinfo {author} {\bibfnamefont
  {A.}~\bibnamefont {Imamoglu}},\ }\href {\doibase
  10.1126/science.290.5500.2282} {\bibfield  {journal} {\bibinfo  {journal}
  {Science}\ }\textbf {\bibinfo {volume} {290}},\ \bibinfo {pages} {2282}
  (\bibinfo {year} {2000})}\BibitemShut {NoStop}%
\bibitem [{\citenamefont {Santori}\ \emph {et~al.}(2001)\citenamefont
  {Santori}, \citenamefont {Pelton}, \citenamefont {Solomon}, \citenamefont
  {Dale},\ and\ \citenamefont {Yamamoto}}]{Santori2001}%
  \BibitemOpen
  \bibfield  {author} {\bibinfo {author} {\bibfnamefont {C.}~\bibnamefont
  {Santori}}, \bibinfo {author} {\bibfnamefont {M.}~\bibnamefont {Pelton}},
  \bibinfo {author} {\bibfnamefont {G.}~\bibnamefont {Solomon}}, \bibinfo
  {author} {\bibfnamefont {Y.}~\bibnamefont {Dale}}, \ and\ \bibinfo {author}
  {\bibfnamefont {Y.}~\bibnamefont {Yamamoto}},\ }\href {\doibase
  10.1103/PhysRevLett.86.1502} {\bibfield  {journal} {\bibinfo  {journal}
  {Phys. Rev. Lett.}\ }\textbf {\bibinfo {volume} {86}},\ \bibinfo {pages}
  {1502} (\bibinfo {year} {2001})}\BibitemShut {NoStop}%
\bibitem [{\citenamefont {He}\ \emph {et~al.}(2013)\citenamefont {He},
  \citenamefont {He}, \citenamefont {Wei}, \citenamefont {Wu}, \citenamefont
  {Atat{\"u}re}, \citenamefont {Schneider}, \citenamefont {H{\"o}fling},
  \citenamefont {Kamp}, \citenamefont {Lu},\ and\ \citenamefont
  {Pan}}]{He2013}%
  \BibitemOpen
  \bibfield  {author} {\bibinfo {author} {\bibfnamefont {Y.-M.}\ \bibnamefont
  {He}}, \bibinfo {author} {\bibfnamefont {Y.}~\bibnamefont {He}}, \bibinfo
  {author} {\bibfnamefont {Y.-J.}\ \bibnamefont {Wei}}, \bibinfo {author}
  {\bibfnamefont {D.}~\bibnamefont {Wu}}, \bibinfo {author} {\bibfnamefont
  {M.}~\bibnamefont {Atat{\"u}re}}, \bibinfo {author} {\bibfnamefont
  {C.}~\bibnamefont {Schneider}}, \bibinfo {author} {\bibfnamefont
  {S.}~\bibnamefont {H{\"o}fling}}, \bibinfo {author} {\bibfnamefont
  {M.}~\bibnamefont {Kamp}}, \bibinfo {author} {\bibfnamefont {C.-Y.}\
  \bibnamefont {Lu}}, \ and\ \bibinfo {author} {\bibfnamefont {J.-W.}\
  \bibnamefont {Pan}},\ }\href {\doibase 10.1038/nnano.2012.262} {\bibfield
  {journal} {\bibinfo  {journal} {Nat. Nanotech.}\ }\textbf {\bibinfo {volume}
  {8}},\ \bibinfo {pages} {213} (\bibinfo {year} {2013})}\BibitemShut {NoStop}%
\bibitem [{\citenamefont {Benson}\ \emph {et~al.}(2000)\citenamefont {Benson},
  \citenamefont {Santori}, \citenamefont {Pelton},\ and\ \citenamefont
  {Yamamoto}}]{Benson2000}%
  \BibitemOpen
  \bibfield  {author} {\bibinfo {author} {\bibfnamefont {O.}~\bibnamefont
  {Benson}}, \bibinfo {author} {\bibfnamefont {C.}~\bibnamefont {Santori}},
  \bibinfo {author} {\bibfnamefont {M.}~\bibnamefont {Pelton}}, \ and\ \bibinfo
  {author} {\bibfnamefont {Y.}~\bibnamefont {Yamamoto}},\ }\href {\doibase
  10.1103/PhysRevLett.84.2513} {\bibfield  {journal} {\bibinfo  {journal}
  {Phys. Rev. Lett.}\ }\textbf {\bibinfo {volume} {84}},\ \bibinfo {pages}
  {2513} (\bibinfo {year} {2000})}\BibitemShut {NoStop}%
\bibitem [{\citenamefont {Stevenson}\ \emph {et~al.}(2006)\citenamefont
  {Stevenson}, \citenamefont {Young}, \citenamefont {Atkinson}, \citenamefont
  {Cooper}, \citenamefont {Ritchie},\ and\ \citenamefont
  {Shields}}]{Stevenson2006}%
  \BibitemOpen
  \bibfield  {author} {\bibinfo {author} {\bibfnamefont {R.~M.}\ \bibnamefont
  {Stevenson}}, \bibinfo {author} {\bibfnamefont {R.~J.}\ \bibnamefont
  {Young}}, \bibinfo {author} {\bibfnamefont {P.}~\bibnamefont {Atkinson}},
  \bibinfo {author} {\bibfnamefont {K.}~\bibnamefont {Cooper}}, \bibinfo
  {author} {\bibfnamefont {D.~A.}\ \bibnamefont {Ritchie}}, \ and\ \bibinfo
  {author} {\bibfnamefont {A.~J.}\ \bibnamefont {Shields}},\ }\href {\doibase
  10.1038/nature04446} {\bibfield  {journal} {\bibinfo  {journal} {Nature}\
  }\textbf {\bibinfo {volume} {439}},\ \bibinfo {pages} {179} (\bibinfo {year}
  {2006})}\BibitemShut {NoStop}%
\bibitem [{\citenamefont {Akopian}\ \emph {et~al.}(2006)\citenamefont
  {Akopian}, \citenamefont {Lindner}, \citenamefont {Poem}, \citenamefont
  {Berlatzky}, \citenamefont {Avron}, \citenamefont {Gershoni}, \citenamefont
  {Gerardot},\ and\ \citenamefont {Petroff}}]{Akopian2006}%
  \BibitemOpen
  \bibfield  {author} {\bibinfo {author} {\bibfnamefont {N.}~\bibnamefont
  {Akopian}}, \bibinfo {author} {\bibfnamefont {N.~H.}\ \bibnamefont
  {Lindner}}, \bibinfo {author} {\bibfnamefont {E.}~\bibnamefont {Poem}},
  \bibinfo {author} {\bibfnamefont {Y.}~\bibnamefont {Berlatzky}}, \bibinfo
  {author} {\bibfnamefont {J.}~\bibnamefont {Avron}}, \bibinfo {author}
  {\bibfnamefont {D.}~\bibnamefont {Gershoni}}, \bibinfo {author}
  {\bibfnamefont {B.~D.}\ \bibnamefont {Gerardot}}, \ and\ \bibinfo {author}
  {\bibfnamefont {P.~M.}\ \bibnamefont {Petroff}},\ }\href {\doibase
  10.1103/PhysRevLett.96.130501} {\bibfield  {journal} {\bibinfo  {journal}
  {Phys. Rev. Lett.}\ }\textbf {\bibinfo {volume} {96}},\ \bibinfo {pages}
  {130501} (\bibinfo {year} {2006})}\BibitemShut {NoStop}%
\bibitem [{\citenamefont {Hafenbrak}\ \emph {et~al.}(2007)\citenamefont
  {Hafenbrak}, \citenamefont {Ulrich}, \citenamefont {Michler}, \citenamefont
  {Wang}, \citenamefont {Rastelli},\ and\ \citenamefont
  {Schmidt}}]{Hafenbrak2007}%
  \BibitemOpen
  \bibfield  {author} {\bibinfo {author} {\bibfnamefont {R.}~\bibnamefont
  {Hafenbrak}}, \bibinfo {author} {\bibfnamefont {S.~M.}\ \bibnamefont
  {Ulrich}}, \bibinfo {author} {\bibfnamefont {P.}~\bibnamefont {Michler}},
  \bibinfo {author} {\bibfnamefont {L.}~\bibnamefont {Wang}}, \bibinfo {author}
  {\bibfnamefont {A.}~\bibnamefont {Rastelli}}, \ and\ \bibinfo {author}
  {\bibfnamefont {O.~G.}\ \bibnamefont {Schmidt}},\ }\href
  {http://stacks.iop.org/1367-2630/9/i=9/a=315} {\bibfield  {journal} {\bibinfo
   {journal} {New J. Phys.}\ }\textbf {\bibinfo {volume} {9}},\ \bibinfo
  {pages} {315} (\bibinfo {year} {2007})}\BibitemShut {NoStop}%
\bibitem [{\citenamefont {Muller}\ \emph {et~al.}(2009)\citenamefont {Muller},
  \citenamefont {Fang}, \citenamefont {Lawall},\ and\ \citenamefont
  {Solomon}}]{Muller2009}%
  \BibitemOpen
  \bibfield  {author} {\bibinfo {author} {\bibfnamefont {A.}~\bibnamefont
  {Muller}}, \bibinfo {author} {\bibfnamefont {W.}~\bibnamefont {Fang}},
  \bibinfo {author} {\bibfnamefont {J.}~\bibnamefont {Lawall}}, \ and\ \bibinfo
  {author} {\bibfnamefont {G.~S.}\ \bibnamefont {Solomon}},\ }\href {\doibase
  10.1103/PhysRevLett.103.217402} {\bibfield  {journal} {\bibinfo  {journal}
  {Phys. Rev. Lett.}\ }\textbf {\bibinfo {volume} {103}},\ \bibinfo {pages}
  {217402} (\bibinfo {year} {2009})}\BibitemShut {NoStop}%
\bibitem [{\citenamefont {Dousse}\ \emph {et~al.}(2010)\citenamefont {Dousse},
  \citenamefont {Suffczy\'{n}ski}, \citenamefont {Beveratos}, \citenamefont
  {Krebs}, \citenamefont {Lema\^{i}tre}, \citenamefont {Sagnes}, \citenamefont
  {Bloch}, \citenamefont {Voisin},\ and\ \citenamefont
  {Senellart}}]{Dousse2010}%
  \BibitemOpen
  \bibfield  {author} {\bibinfo {author} {\bibfnamefont {A.}~\bibnamefont
  {Dousse}}, \bibinfo {author} {\bibfnamefont {J.}~\bibnamefont
  {Suffczy\'{n}ski}}, \bibinfo {author} {\bibfnamefont {A.}~\bibnamefont
  {Beveratos}}, \bibinfo {author} {\bibfnamefont {O.}~\bibnamefont {Krebs}},
  \bibinfo {author} {\bibfnamefont {A.}~\bibnamefont {Lema\^{i}tre}}, \bibinfo
  {author} {\bibfnamefont {I.}~\bibnamefont {Sagnes}}, \bibinfo {author}
  {\bibfnamefont {J.}~\bibnamefont {Bloch}}, \bibinfo {author} {\bibfnamefont
  {P.}~\bibnamefont {Voisin}}, \ and\ \bibinfo {author} {\bibfnamefont
  {P.}~\bibnamefont {Senellart}},\ }\href {\doibase 10.1038/nature09148}
  {\bibfield  {journal} {\bibinfo  {journal} {Nature}\ }\textbf {\bibinfo
  {volume} {466}},\ \bibinfo {pages} {217} (\bibinfo {year}
  {2010})}\BibitemShut {NoStop}%
\bibitem [{\citenamefont {Somaschi}\ \emph {et~al.}(2016)\citenamefont
  {Somaschi}, \citenamefont {Giesz}, \citenamefont {De~Santis}, \citenamefont
  {Loredo}, \citenamefont {Almeida}, \citenamefont {Hornecker}, \citenamefont
  {Portalupi}, \citenamefont {Grange}, \citenamefont {Ant\'{o}n}, \citenamefont
  {Demory}, \citenamefont {G\'{o}mez}, \citenamefont {Sagnes}, \citenamefont
  {Lanzillotti-Kimura}, \citenamefont {Lema\'{i}tre}, \citenamefont {Auffeves},
  \citenamefont {White}, \citenamefont {Lanco},\ and\ \citenamefont
  {Senellart}}]{Somaschi2016}%
  \BibitemOpen
  \bibfield  {author} {\bibinfo {author} {\bibfnamefont {N.}~\bibnamefont
  {Somaschi}}, \bibinfo {author} {\bibfnamefont {V.}~\bibnamefont {Giesz}},
  \bibinfo {author} {\bibfnamefont {L.}~\bibnamefont {De~Santis}}, \bibinfo
  {author} {\bibfnamefont {J.~C.}\ \bibnamefont {Loredo}}, \bibinfo {author}
  {\bibfnamefont {M.~P.}\ \bibnamefont {Almeida}}, \bibinfo {author}
  {\bibfnamefont {G.}~\bibnamefont {Hornecker}}, \bibinfo {author}
  {\bibfnamefont {S.~L.}\ \bibnamefont {Portalupi}}, \bibinfo {author}
  {\bibfnamefont {T.}~\bibnamefont {Grange}}, \bibinfo {author} {\bibfnamefont
  {C.}~\bibnamefont {Ant\'{o}n}}, \bibinfo {author} {\bibfnamefont
  {J.}~\bibnamefont {Demory}}, \bibinfo {author} {\bibfnamefont
  {C.}~\bibnamefont {G\'{o}mez}}, \bibinfo {author} {\bibfnamefont
  {I.}~\bibnamefont {Sagnes}}, \bibinfo {author} {\bibfnamefont {N.~D.}\
  \bibnamefont {Lanzillotti-Kimura}}, \bibinfo {author} {\bibfnamefont
  {A.}~\bibnamefont {Lema\'{i}tre}}, \bibinfo {author} {\bibfnamefont
  {A.}~\bibnamefont {Auffeves}}, \bibinfo {author} {\bibfnamefont {A.~G.}\
  \bibnamefont {White}}, \bibinfo {author} {\bibfnamefont {L.}~\bibnamefont
  {Lanco}}, \ and\ \bibinfo {author} {\bibfnamefont {P.}~\bibnamefont
  {Senellart}},\ }\href {\doibase 10.1038/nphoton.2016.23} {\bibfield
  {journal} {\bibinfo  {journal} {Nat. Photon.}\ }\textbf {\bibinfo {volume}
  {10}},\ \bibinfo {pages} {340} (\bibinfo {year} {2016})}\BibitemShut
  {NoStop}%
\bibitem [{\citenamefont {Ding}\ \emph {et~al.}(2016)\citenamefont {Ding},
  \citenamefont {He}, \citenamefont {Duan}, \citenamefont {Gregersen},
  \citenamefont {Chen}, \citenamefont {Unsleber}, \citenamefont {Maier},
  \citenamefont {Schneider}, \citenamefont {Kamp}, \citenamefont {H\"ofling},
  \citenamefont {Lu},\ and\ \citenamefont {Pan}}]{Ding2016}%
  \BibitemOpen
  \bibfield  {author} {\bibinfo {author} {\bibfnamefont {X.}~\bibnamefont
  {Ding}}, \bibinfo {author} {\bibfnamefont {Y.}~\bibnamefont {He}}, \bibinfo
  {author} {\bibfnamefont {Z.-C.}\ \bibnamefont {Duan}}, \bibinfo {author}
  {\bibfnamefont {N.}~\bibnamefont {Gregersen}}, \bibinfo {author}
  {\bibfnamefont {M.-C.}\ \bibnamefont {Chen}}, \bibinfo {author}
  {\bibfnamefont {S.}~\bibnamefont {Unsleber}}, \bibinfo {author}
  {\bibfnamefont {S.}~\bibnamefont {Maier}}, \bibinfo {author} {\bibfnamefont
  {C.}~\bibnamefont {Schneider}}, \bibinfo {author} {\bibfnamefont
  {M.}~\bibnamefont {Kamp}}, \bibinfo {author} {\bibfnamefont {S.}~\bibnamefont
  {H\"ofling}}, \bibinfo {author} {\bibfnamefont {C.-Y.}\ \bibnamefont {Lu}}, \
  and\ \bibinfo {author} {\bibfnamefont {J.-W.}\ \bibnamefont {Pan}},\ }\href
  {\doibase 10.1103/PhysRevLett.116.020401} {\bibfield  {journal} {\bibinfo
  {journal} {Phys. Rev. Lett.}\ }\textbf {\bibinfo {volume} {116}},\ \bibinfo
  {pages} {020401} (\bibinfo {year} {2016})}\BibitemShut {NoStop}%
\bibitem [{\citenamefont {Keil}\ \emph {et~al.}(2017)\citenamefont {Keil},
  \citenamefont {Zopf}, \citenamefont {Chen}, \citenamefont {H{\"o}fer},
  \citenamefont {Zhang}, \citenamefont {Ding},\ and\ \citenamefont
  {Schmidt}}]{Keil2017}%
  \BibitemOpen
  \bibfield  {author} {\bibinfo {author} {\bibfnamefont {R.}~\bibnamefont
  {Keil}}, \bibinfo {author} {\bibfnamefont {M.}~\bibnamefont {Zopf}}, \bibinfo
  {author} {\bibfnamefont {Y.}~\bibnamefont {Chen}}, \bibinfo {author}
  {\bibfnamefont {B.}~\bibnamefont {H{\"o}fer}}, \bibinfo {author}
  {\bibfnamefont {J.}~\bibnamefont {Zhang}}, \bibinfo {author} {\bibfnamefont
  {F.}~\bibnamefont {Ding}}, \ and\ \bibinfo {author} {\bibfnamefont {O.~G.}\
  \bibnamefont {Schmidt}},\ }\href {http://dx.doi.org/10.1038/ncomms15501}
  {\bibfield  {journal} {\bibinfo  {journal} {Nat. Commun.}\ }\textbf {\bibinfo
  {volume} {8}},\ \bibinfo {pages} {15501} (\bibinfo {year}
  {2017})}\BibitemShut {NoStop}%
\bibitem [{\citenamefont {Atkinson}\ \emph {et~al.}(2012)\citenamefont
  {Atkinson}, \citenamefont {Zallo},\ and\ \citenamefont
  {Schmidt}}]{Atkinson2012}%
  \BibitemOpen
  \bibfield  {author} {\bibinfo {author} {\bibfnamefont {P.}~\bibnamefont
  {Atkinson}}, \bibinfo {author} {\bibfnamefont {E.}~\bibnamefont {Zallo}}, \
  and\ \bibinfo {author} {\bibfnamefont {O.~G.}\ \bibnamefont {Schmidt}},\
  }\href {\doibase 10.1063/1.4748183} {\bibfield  {journal} {\bibinfo
  {journal} {J. Appl. Phys.}\ }\textbf {\bibinfo {volume} {112}},\ \bibinfo
  {pages} {054303} (\bibinfo {year} {2012})}\BibitemShut {NoStop}%
\bibitem [{\citenamefont {Kumar}\ \emph {et~al.}(2011)\citenamefont {Kumar},
  \citenamefont {Trotta}, \citenamefont {Zallo}, \citenamefont {Plumhof},
  \citenamefont {Atkinson}, \citenamefont {Rastelli},\ and\ \citenamefont
  {Schmidt}}]{Kumar2011}%
  \BibitemOpen
  \bibfield  {author} {\bibinfo {author} {\bibfnamefont {S.}~\bibnamefont
  {Kumar}}, \bibinfo {author} {\bibfnamefont {R.}~\bibnamefont {Trotta}},
  \bibinfo {author} {\bibfnamefont {E.}~\bibnamefont {Zallo}}, \bibinfo
  {author} {\bibfnamefont {J.~D.}\ \bibnamefont {Plumhof}}, \bibinfo {author}
  {\bibfnamefont {P.}~\bibnamefont {Atkinson}}, \bibinfo {author}
  {\bibfnamefont {A.}~\bibnamefont {Rastelli}}, \ and\ \bibinfo {author}
  {\bibfnamefont {O.~G.}\ \bibnamefont {Schmidt}},\ }\href {\doibase
  10.1063/1.3653804} {\bibfield  {journal} {\bibinfo  {journal} {Appl. Phys.
  Lett.}\ }\textbf {\bibinfo {volume} {99}},\ \bibinfo {pages} {161118}
  (\bibinfo {year} {2011})}\BibitemShut {NoStop}%
\bibitem [{\citenamefont {Jahn}\ \emph {et~al.}(2015)\citenamefont {Jahn},
  \citenamefont {Munsch}, \citenamefont {B\'eguin}, \citenamefont {Kuhlmann},
  \citenamefont {Renggli}, \citenamefont {Huo}, \citenamefont {Ding},
  \citenamefont {Trotta}, \citenamefont {Reindl}, \citenamefont {Schmidt},
  \citenamefont {Rastelli}, \citenamefont {Treutlein},\ and\ \citenamefont
  {Warburton}}]{Jahn2015}%
  \BibitemOpen
  \bibfield  {author} {\bibinfo {author} {\bibfnamefont {J.-P.}\ \bibnamefont
  {Jahn}}, \bibinfo {author} {\bibfnamefont {M.}~\bibnamefont {Munsch}},
  \bibinfo {author} {\bibfnamefont {L.}~\bibnamefont {B\'eguin}}, \bibinfo
  {author} {\bibfnamefont {A.~V.}\ \bibnamefont {Kuhlmann}}, \bibinfo {author}
  {\bibfnamefont {M.}~\bibnamefont {Renggli}}, \bibinfo {author} {\bibfnamefont
  {Y.}~\bibnamefont {Huo}}, \bibinfo {author} {\bibfnamefont {F.}~\bibnamefont
  {Ding}}, \bibinfo {author} {\bibfnamefont {R.}~\bibnamefont {Trotta}},
  \bibinfo {author} {\bibfnamefont {M.}~\bibnamefont {Reindl}}, \bibinfo
  {author} {\bibfnamefont {O.~G.}\ \bibnamefont {Schmidt}}, \bibinfo {author}
  {\bibfnamefont {A.}~\bibnamefont {Rastelli}}, \bibinfo {author}
  {\bibfnamefont {P.}~\bibnamefont {Treutlein}}, \ and\ \bibinfo {author}
  {\bibfnamefont {R.~J.}\ \bibnamefont {Warburton}},\ }\href {\doibase
  10.1103/PhysRevB.92.245439} {\bibfield  {journal} {\bibinfo  {journal} {Phys.
  Rev. B}\ }\textbf {\bibinfo {volume} {92}},\ \bibinfo {pages} {245439}
  (\bibinfo {year} {2015})}\BibitemShut {NoStop}%
\bibitem [{\citenamefont {Giesz}\ \emph {et~al.}(2015)\citenamefont {Giesz},
  \citenamefont {Portalupi}, \citenamefont {Grange}, \citenamefont {Ant\'{o}n},
  \citenamefont {De~Santis}, \citenamefont {Demory}, \citenamefont {Somaschi},
  \citenamefont {Sagnes}, \citenamefont {Lema\^{i}tre}, \citenamefont {Lanco},
  \citenamefont {Auff\`{e}ves},\ and\ \citenamefont {Senellart}}]{Giesz2015}%
  \BibitemOpen
  \bibfield  {author} {\bibinfo {author} {\bibfnamefont {V.}~\bibnamefont
  {Giesz}}, \bibinfo {author} {\bibfnamefont {S.~L.}\ \bibnamefont
  {Portalupi}}, \bibinfo {author} {\bibfnamefont {T.}~\bibnamefont {Grange}},
  \bibinfo {author} {\bibfnamefont {C.}~\bibnamefont {Ant\'{o}n}}, \bibinfo
  {author} {\bibfnamefont {L.}~\bibnamefont {De~Santis}}, \bibinfo {author}
  {\bibfnamefont {J.}~\bibnamefont {Demory}}, \bibinfo {author} {\bibfnamefont
  {N.}~\bibnamefont {Somaschi}}, \bibinfo {author} {\bibfnamefont
  {I.}~\bibnamefont {Sagnes}}, \bibinfo {author} {\bibfnamefont
  {A.}~\bibnamefont {Lema\^{i}tre}}, \bibinfo {author} {\bibfnamefont
  {L.}~\bibnamefont {Lanco}}, \bibinfo {author} {\bibfnamefont
  {A.}~\bibnamefont {Auff\`{e}ves}}, \ and\ \bibinfo {author} {\bibfnamefont
  {P.}~\bibnamefont {Senellart}},\ }\href {\doibase 10.1103/PhysRevB.92.161302}
  {\bibfield  {journal} {\bibinfo  {journal} {Phys. Rev. B}\ }\textbf {\bibinfo
  {volume} {92}},\ \bibinfo {pages} {161302} (\bibinfo {year}
  {2015})}\BibitemShut {NoStop}%
\bibitem [{\citenamefont {Bennett}\ \emph {et~al.}(2010)\citenamefont
  {Bennett}, \citenamefont {Pooley}, \citenamefont {Stevenson}, \citenamefont
  {Ward}, \citenamefont {Patel}, \citenamefont {de~la Giroday}, \citenamefont
  {Sk\"{o}ld}, \citenamefont {Farrer}, \citenamefont {Nicoll}, \citenamefont
  {Ritchie},\ and\ \citenamefont {Shields}}]{Bennett2010}%
  \BibitemOpen
  \bibfield  {author} {\bibinfo {author} {\bibfnamefont {A.~J.}\ \bibnamefont
  {Bennett}}, \bibinfo {author} {\bibfnamefont {M.~A.}\ \bibnamefont {Pooley}},
  \bibinfo {author} {\bibfnamefont {R.~M.}\ \bibnamefont {Stevenson}}, \bibinfo
  {author} {\bibfnamefont {M.~B.}\ \bibnamefont {Ward}}, \bibinfo {author}
  {\bibfnamefont {R.~B.}\ \bibnamefont {Patel}}, \bibinfo {author}
  {\bibfnamefont {A.~B.}\ \bibnamefont {de~la Giroday}}, \bibinfo {author}
  {\bibfnamefont {N.}~\bibnamefont {Sk\"{o}ld}}, \bibinfo {author}
  {\bibfnamefont {I.}~\bibnamefont {Farrer}}, \bibinfo {author} {\bibfnamefont
  {C.~A.}\ \bibnamefont {Nicoll}}, \bibinfo {author} {\bibfnamefont {D.~A.}\
  \bibnamefont {Ritchie}}, \ and\ \bibinfo {author} {\bibfnamefont {A.~J.}\
  \bibnamefont {Shields}},\ }\href {\doibase 10.1038/nphys1780} {\bibfield
  {journal} {\bibinfo  {journal} {Nat. Phys.}\ }\textbf {\bibinfo {volume}
  {6}},\ \bibinfo {pages} {947} (\bibinfo {year} {2010})}\BibitemShut {NoStop}%
\bibitem [{\citenamefont {Ghali}\ \emph {et~al.}(2012)\citenamefont {Ghali},
  \citenamefont {Ohtani}, \citenamefont {Ohno},\ and\ \citenamefont
  {Ohno}}]{Ghali2012}%
  \BibitemOpen
  \bibfield  {author} {\bibinfo {author} {\bibfnamefont {M.}~\bibnamefont
  {Ghali}}, \bibinfo {author} {\bibfnamefont {K.}~\bibnamefont {Ohtani}},
  \bibinfo {author} {\bibfnamefont {Y.}~\bibnamefont {Ohno}}, \ and\ \bibinfo
  {author} {\bibfnamefont {H.}~\bibnamefont {Ohno}},\ }\href {\doibase
  10.1038/ncomms1657} {\bibfield  {journal} {\bibinfo  {journal} {Nat.
  Commun.}\ }\textbf {\bibinfo {volume} {3}},\ \bibinfo {pages} {661} (\bibinfo
  {year} {2012})}\BibitemShut {NoStop}%
\bibitem [{\citenamefont {Zhang}\ \emph {et~al.}(2017)\citenamefont {Zhang},
  \citenamefont {Zallo}, \citenamefont {H\"{o}fer}, \citenamefont {Chen},
  \citenamefont {Keil}, \citenamefont {Zopf}, \citenamefont {B\"{o}ttner},
  \citenamefont {Ding},\ and\ \citenamefont {Schmidt}}]{Zhang2017}%
  \BibitemOpen
  \bibfield  {author} {\bibinfo {author} {\bibfnamefont {J.}~\bibnamefont
  {Zhang}}, \bibinfo {author} {\bibfnamefont {E.}~\bibnamefont {Zallo}},
  \bibinfo {author} {\bibfnamefont {B.}~\bibnamefont {H\"{o}fer}}, \bibinfo
  {author} {\bibfnamefont {Y.}~\bibnamefont {Chen}}, \bibinfo {author}
  {\bibfnamefont {R.}~\bibnamefont {Keil}}, \bibinfo {author} {\bibfnamefont
  {M.}~\bibnamefont {Zopf}}, \bibinfo {author} {\bibfnamefont {S.}~\bibnamefont
  {B\"{o}ttner}}, \bibinfo {author} {\bibfnamefont {F.}~\bibnamefont {Ding}}, \
  and\ \bibinfo {author} {\bibfnamefont {O.~G.}\ \bibnamefont {Schmidt}},\
  }\href {\doibase 10.1021/acs.nanolett.6b04539} {\bibfield  {journal}
  {\bibinfo  {journal} {Nano Lett.}\ }\textbf {\bibinfo {volume} {17}},\
  \bibinfo {pages} {501} (\bibinfo {year} {2017})}\BibitemShut {NoStop}%
\bibitem [{\citenamefont {Pooley}\ \emph {et~al.}(2014)\citenamefont {Pooley},
  \citenamefont {Bennett}, \citenamefont {Stevenson}, \citenamefont {Shields},
  \citenamefont {Farrer},\ and\ \citenamefont {Ritchie}}]{Pooley2014}%
  \BibitemOpen
  \bibfield  {author} {\bibinfo {author} {\bibfnamefont {M.~A.}\ \bibnamefont
  {Pooley}}, \bibinfo {author} {\bibfnamefont {A.~J.}\ \bibnamefont {Bennett}},
  \bibinfo {author} {\bibfnamefont {R.~M.}\ \bibnamefont {Stevenson}}, \bibinfo
  {author} {\bibfnamefont {A.~J.}\ \bibnamefont {Shields}}, \bibinfo {author}
  {\bibfnamefont {I.}~\bibnamefont {Farrer}}, \ and\ \bibinfo {author}
  {\bibfnamefont {D.~A.}\ \bibnamefont {Ritchie}},\ }\href {\doibase
  10.1103/PhysRevApplied.1.024002} {\bibfield  {journal} {\bibinfo  {journal}
  {Phys. Rev. Applied}\ }\textbf {\bibinfo {volume} {1}},\ \bibinfo {pages}
  {024002} (\bibinfo {year} {2014})}\BibitemShut {NoStop}%
\bibitem [{\citenamefont {Ding}\ \emph {et~al.}(2010)\citenamefont {Ding},
  \citenamefont {Singh}, \citenamefont {Plumhof}, \citenamefont {Zander},
  \citenamefont {K\ifmmode~\check{r}\else \v{r}\fi{}\'apek}, \citenamefont
  {Chen}, \citenamefont {Benyoucef}, \citenamefont {Zwiller}, \citenamefont
  {D\"orr}, \citenamefont {Bester}, \citenamefont {Rastelli},\ and\
  \citenamefont {Schmidt}}]{Ding2010a}%
  \BibitemOpen
  \bibfield  {author} {\bibinfo {author} {\bibfnamefont {F.}~\bibnamefont
  {Ding}}, \bibinfo {author} {\bibfnamefont {R.}~\bibnamefont {Singh}},
  \bibinfo {author} {\bibfnamefont {J.~D.}\ \bibnamefont {Plumhof}}, \bibinfo
  {author} {\bibfnamefont {T.}~\bibnamefont {Zander}}, \bibinfo {author}
  {\bibfnamefont {V.}~\bibnamefont {K\ifmmode~\check{r}\else
  \v{r}\fi{}\'apek}}, \bibinfo {author} {\bibfnamefont {Y.~H.}\ \bibnamefont
  {Chen}}, \bibinfo {author} {\bibfnamefont {M.}~\bibnamefont {Benyoucef}},
  \bibinfo {author} {\bibfnamefont {V.}~\bibnamefont {Zwiller}}, \bibinfo
  {author} {\bibfnamefont {K.}~\bibnamefont {D\"orr}}, \bibinfo {author}
  {\bibfnamefont {G.}~\bibnamefont {Bester}}, \bibinfo {author} {\bibfnamefont
  {A.}~\bibnamefont {Rastelli}}, \ and\ \bibinfo {author} {\bibfnamefont
  {O.~G.}\ \bibnamefont {Schmidt}},\ }\href {\doibase
  10.1103/PhysRevLett.104.067405} {\bibfield  {journal} {\bibinfo  {journal}
  {Phys. Rev. Lett.}\ }\textbf {\bibinfo {volume} {104}},\ \bibinfo {pages}
  {067405} (\bibinfo {year} {2010})}\BibitemShut {NoStop}%
\bibitem [{\citenamefont {Zhang}\ \emph {et~al.}(2015)\citenamefont {Zhang},
  \citenamefont {Huo}, \citenamefont {Rastelli}, \citenamefont {Zopf},
  \citenamefont {H\"{o}fer}, \citenamefont {Chen}, \citenamefont {Ding},\ and\
  \citenamefont {Schmidt}}]{Zhang2015}%
  \BibitemOpen
  \bibfield  {author} {\bibinfo {author} {\bibfnamefont {J.}~\bibnamefont
  {Zhang}}, \bibinfo {author} {\bibfnamefont {Y.}~\bibnamefont {Huo}}, \bibinfo
  {author} {\bibfnamefont {A.}~\bibnamefont {Rastelli}}, \bibinfo {author}
  {\bibfnamefont {M.}~\bibnamefont {Zopf}}, \bibinfo {author} {\bibfnamefont
  {B.}~\bibnamefont {H\"{o}fer}}, \bibinfo {author} {\bibfnamefont
  {Y.}~\bibnamefont {Chen}}, \bibinfo {author} {\bibfnamefont {F.}~\bibnamefont
  {Ding}}, \ and\ \bibinfo {author} {\bibfnamefont {O.~G.}\ \bibnamefont
  {Schmidt}},\ }\href {\doibase 10.1021/nl5037512} {\bibfield  {journal}
  {\bibinfo  {journal} {Nano Lett.}\ }\textbf {\bibinfo {volume} {15}},\
  \bibinfo {pages} {422} (\bibinfo {year} {2015})}\BibitemShut {NoStop}%
\bibitem [{\citenamefont {Chen}\ \emph {et~al.}(2016)\citenamefont {Chen},
  \citenamefont {Zhang}, \citenamefont {Zopf}, \citenamefont {Jung},
  \citenamefont {Zhang}, \citenamefont {Keil}, \citenamefont {Ding},\ and\
  \citenamefont {Schmidt}}]{Chen2016}%
  \BibitemOpen
  \bibfield  {author} {\bibinfo {author} {\bibfnamefont {Y.}~\bibnamefont
  {Chen}}, \bibinfo {author} {\bibfnamefont {J.}~\bibnamefont {Zhang}},
  \bibinfo {author} {\bibfnamefont {M.}~\bibnamefont {Zopf}}, \bibinfo {author}
  {\bibfnamefont {K.}~\bibnamefont {Jung}}, \bibinfo {author} {\bibfnamefont
  {Y.}~\bibnamefont {Zhang}}, \bibinfo {author} {\bibfnamefont
  {R.}~\bibnamefont {Keil}}, \bibinfo {author} {\bibfnamefont {F.}~\bibnamefont
  {Ding}}, \ and\ \bibinfo {author} {\bibfnamefont {O.~G.}\ \bibnamefont
  {Schmidt}},\ }\href {\doibase 10.1038/ncomms10387} {\bibfield  {journal}
  {\bibinfo  {journal} {Nat. Commun.}\ }\textbf {\bibinfo {volume} {7}},\
  \bibinfo {pages} {10387} (\bibinfo {year} {2016})}\BibitemShut {NoStop}%
\bibitem [{\citenamefont {Metcalfe}\ \emph {et~al.}(2009)\citenamefont
  {Metcalfe}, \citenamefont {Muller}, \citenamefont {Solomon},\ and\
  \citenamefont {Lawall}}]{Metcalfe2009}%
  \BibitemOpen
  \bibfield  {author} {\bibinfo {author} {\bibfnamefont {M.}~\bibnamefont
  {Metcalfe}}, \bibinfo {author} {\bibfnamefont {A.}~\bibnamefont {Muller}},
  \bibinfo {author} {\bibfnamefont {G.~S.}\ \bibnamefont {Solomon}}, \ and\
  \bibinfo {author} {\bibfnamefont {J.}~\bibnamefont {Lawall}},\ }\href
  {\doibase 10.1364/JOSAB.26.002308} {\bibfield  {journal} {\bibinfo  {journal}
  {J. Opt. Soc. Am. B}\ }\textbf {\bibinfo {volume} {26}},\ \bibinfo {pages}
  {2308} (\bibinfo {year} {2009})}\BibitemShut {NoStop}%
\bibitem [{\citenamefont {Prechtel}\ \emph {et~al.}(2013)\citenamefont
  {Prechtel}, \citenamefont {Kuhlmann}, \citenamefont {Houel}, \citenamefont
  {Greuter}, \citenamefont {Ludwig}, \citenamefont {Reuter}, \citenamefont
  {Wieck},\ and\ \citenamefont {Warburton}}]{Prechtel2013}%
  \BibitemOpen
  \bibfield  {author} {\bibinfo {author} {\bibfnamefont {J.~H.}\ \bibnamefont
  {Prechtel}}, \bibinfo {author} {\bibfnamefont {A.~V.}\ \bibnamefont
  {Kuhlmann}}, \bibinfo {author} {\bibfnamefont {J.}~\bibnamefont {Houel}},
  \bibinfo {author} {\bibfnamefont {L.}~\bibnamefont {Greuter}}, \bibinfo
  {author} {\bibfnamefont {A.}~\bibnamefont {Ludwig}}, \bibinfo {author}
  {\bibfnamefont {D.}~\bibnamefont {Reuter}}, \bibinfo {author} {\bibfnamefont
  {A.~D.}\ \bibnamefont {Wieck}}, \ and\ \bibinfo {author} {\bibfnamefont
  {R.~J.}\ \bibnamefont {Warburton}},\ }\href {\doibase
  10.1103/PhysRevX.3.041006} {\bibfield  {journal} {\bibinfo  {journal} {Phys.
  Rev. X}\ }\textbf {\bibinfo {volume} {3}},\ \bibinfo {pages} {041006}
  (\bibinfo {year} {2013})}\BibitemShut {NoStop}%
\bibitem [{\citenamefont {Akopian}\ \emph {et~al.}(2013)\citenamefont
  {Akopian}, \citenamefont {Trotta}, \citenamefont {Zallo}, \citenamefont
  {Kumar}, \citenamefont {Atkinson}, \citenamefont {Rastelli}, \citenamefont
  {Schmidt},\ and\ \citenamefont {Zwiller}}]{Akopian2013}%
  \BibitemOpen
  \bibfield  {author} {\bibinfo {author} {\bibfnamefont {N.}~\bibnamefont
  {Akopian}}, \bibinfo {author} {\bibfnamefont {R.}~\bibnamefont {Trotta}},
  \bibinfo {author} {\bibfnamefont {E.}~\bibnamefont {Zallo}}, \bibinfo
  {author} {\bibfnamefont {S.}~\bibnamefont {Kumar}}, \bibinfo {author}
  {\bibfnamefont {P.}~\bibnamefont {Atkinson}}, \bibinfo {author}
  {\bibfnamefont {A.}~\bibnamefont {Rastelli}}, \bibinfo {author}
  {\bibfnamefont {O.~G.}\ \bibnamefont {Schmidt}}, \ and\ \bibinfo {author}
  {\bibfnamefont {V.}~\bibnamefont {Zwiller}},\ }\href
  {http://arxiv.org/abs/1302.2005} {\bibfield  {journal} {\bibinfo  {journal}
  {arXiv preprint: 1302.2005}\ } (\bibinfo {year} {2013})}\BibitemShut
  {NoStop}%
\bibitem [{\citenamefont {Trotta}\ \emph {et~al.}(2012)\citenamefont {Trotta},
  \citenamefont {Atkinson}, \citenamefont {Plumhof}, \citenamefont {Zallo},
  \citenamefont {Rezaev}, \citenamefont {Kumar}, \citenamefont {Baunack},
  \citenamefont {Schr\"{o}ter}, \citenamefont {Rastelli},\ and\ \citenamefont
  {Schmidt}}]{Trotta2012}%
  \BibitemOpen
  \bibfield  {author} {\bibinfo {author} {\bibfnamefont {R.}~\bibnamefont
  {Trotta}}, \bibinfo {author} {\bibfnamefont {P.}~\bibnamefont {Atkinson}},
  \bibinfo {author} {\bibfnamefont {J.~D.}\ \bibnamefont {Plumhof}}, \bibinfo
  {author} {\bibfnamefont {E.}~\bibnamefont {Zallo}}, \bibinfo {author}
  {\bibfnamefont {R.~O.}\ \bibnamefont {Rezaev}}, \bibinfo {author}
  {\bibfnamefont {S.}~\bibnamefont {Kumar}}, \bibinfo {author} {\bibfnamefont
  {S.}~\bibnamefont {Baunack}}, \bibinfo {author} {\bibfnamefont {J.~R.}\
  \bibnamefont {Schr\"{o}ter}}, \bibinfo {author} {\bibfnamefont
  {A.}~\bibnamefont {Rastelli}}, \ and\ \bibinfo {author} {\bibfnamefont
  {O.~G.}\ \bibnamefont {Schmidt}},\ }\href {\doibase 10.1002/adma.201200537}
  {\bibfield  {journal} {\bibinfo  {journal} {Advanced Materials}\ }\textbf
  {\bibinfo {volume} {24}},\ \bibinfo {pages} {2668} (\bibinfo {year}
  {2012})}\BibitemShut {NoStop}%
\bibitem [{\citenamefont {Rakher}\ \emph {et~al.}(2013)\citenamefont {Rakher},
  \citenamefont {Warburton},\ and\ \citenamefont {Treutlein}}]{Rakher2013}%
  \BibitemOpen
  \bibfield  {author} {\bibinfo {author} {\bibfnamefont {M.~T.}\ \bibnamefont
  {Rakher}}, \bibinfo {author} {\bibfnamefont {R.~J.}\ \bibnamefont
  {Warburton}}, \ and\ \bibinfo {author} {\bibfnamefont {P.}~\bibnamefont
  {Treutlein}},\ }\href {\doibase 10.1103/PhysRevA.88.053834} {\bibfield
  {journal} {\bibinfo  {journal} {Phys. Rev. A}\ }\textbf {\bibinfo {volume}
  {88}},\ \bibinfo {pages} {1} (\bibinfo {year} {2013})}\BibitemShut {NoStop}%
\bibitem [{\citenamefont {Wolters}\ \emph {et~al.}(2017)\citenamefont
  {Wolters}, \citenamefont {Buser}, \citenamefont {Horsley}, \citenamefont
  {B{\'{e}}guin}, \citenamefont {J{\"{o}}ckel}, \citenamefont {Jahn},
  \citenamefont {Warburton},\ and\ \citenamefont {Treutlein}}]{Wolters2017}%
  \BibitemOpen
  \bibfield  {author} {\bibinfo {author} {\bibfnamefont {J.}~\bibnamefont
  {Wolters}}, \bibinfo {author} {\bibfnamefont {G.}~\bibnamefont {Buser}},
  \bibinfo {author} {\bibfnamefont {A.}~\bibnamefont {Horsley}}, \bibinfo
  {author} {\bibfnamefont {L.}~\bibnamefont {B{\'{e}}guin}}, \bibinfo {author}
  {\bibfnamefont {A.}~\bibnamefont {J{\"{o}}ckel}}, \bibinfo {author}
  {\bibfnamefont {J.~P.}\ \bibnamefont {Jahn}}, \bibinfo {author}
  {\bibfnamefont {R.~J.}\ \bibnamefont {Warburton}}, \ and\ \bibinfo {author}
  {\bibfnamefont {P.}~\bibnamefont {Treutlein}},\ }\href {\doibase
  10.1103/PhysRevLett.119.060502} {\bibfield  {journal} {\bibinfo  {journal}
  {Phys. Rev. Lett.}\ }\textbf {\bibinfo {volume} {119}},\ \bibinfo {pages} {1}
  (\bibinfo {year} {2017})}\BibitemShut {NoStop}%
\bibitem [{\citenamefont {Hong}\ \emph {et~al.}(1987)\citenamefont {Hong},
  \citenamefont {Ou},\ and\ \citenamefont {Mandel}}]{Hong1987}%
  \BibitemOpen
  \bibfield  {author} {\bibinfo {author} {\bibfnamefont {C.~K.}\ \bibnamefont
  {Hong}}, \bibinfo {author} {\bibfnamefont {Z.~Y.}\ \bibnamefont {Ou}}, \ and\
  \bibinfo {author} {\bibfnamefont {L.}~\bibnamefont {Mandel}},\ }\href
  {\doibase 10.1103/PhysRevLett.59.2044} {\bibfield  {journal} {\bibinfo
  {journal} {Phys. Rev. Lett.}\ }\textbf {\bibinfo {volume} {59}},\ \bibinfo
  {pages} {2044} (\bibinfo {year} {1987})}\BibitemShut {NoStop}%
\bibitem [{\citenamefont {Patel}\ \emph {et~al.}(2010)\citenamefont {Patel},
  \citenamefont {Bennett}, \citenamefont {Farrer}, \citenamefont {Nicoll},
  \citenamefont {Ritchie},\ and\ \citenamefont {Shields}}]{Patel2010}%
  \BibitemOpen
  \bibfield  {author} {\bibinfo {author} {\bibfnamefont {R.~B.}\ \bibnamefont
  {Patel}}, \bibinfo {author} {\bibfnamefont {A.~J.}\ \bibnamefont {Bennett}},
  \bibinfo {author} {\bibfnamefont {I.}~\bibnamefont {Farrer}}, \bibinfo
  {author} {\bibfnamefont {C.~A.}\ \bibnamefont {Nicoll}}, \bibinfo {author}
  {\bibfnamefont {D.~A.}\ \bibnamefont {Ritchie}}, \ and\ \bibinfo {author}
  {\bibfnamefont {A.~J.}\ \bibnamefont {Shields}},\ }\href {\doibase
  10.1038/nphoton.2010.161} {\bibfield  {journal} {\bibinfo  {journal} {Nat.
  Photon.}\ }\textbf {\bibinfo {volume} {4}},\ \bibinfo {pages} {632} (\bibinfo
  {year} {2010})}\BibitemShut {NoStop}%
\bibitem [{\citenamefont {Flagg}\ \emph {et~al.}(2010)\citenamefont {Flagg},
  \citenamefont {Muller}, \citenamefont {Polyakov}, \citenamefont {Ling},
  \citenamefont {Migdall},\ and\ \citenamefont {Solomon}}]{Flagg2010}%
  \BibitemOpen
  \bibfield  {author} {\bibinfo {author} {\bibfnamefont {E.~B.}\ \bibnamefont
  {Flagg}}, \bibinfo {author} {\bibfnamefont {A.}~\bibnamefont {Muller}},
  \bibinfo {author} {\bibfnamefont {S.~V.}\ \bibnamefont {Polyakov}}, \bibinfo
  {author} {\bibfnamefont {A.}~\bibnamefont {Ling}}, \bibinfo {author}
  {\bibfnamefont {A.}~\bibnamefont {Migdall}}, \ and\ \bibinfo {author}
  {\bibfnamefont {G.~S.}\ \bibnamefont {Solomon}},\ }\href {\doibase
  10.1103/PhysRevLett.104.137401} {\bibfield  {journal} {\bibinfo  {journal}
  {Phys. Rev. Lett.}\ }\textbf {\bibinfo {volume} {104}},\ \bibinfo {pages}
  {137401} (\bibinfo {year} {2010})}\BibitemShut {NoStop}%
\bibitem [{\citenamefont {Zhang}\ \emph {et~al.}(2016)\citenamefont {Zhang},
  \citenamefont {Chen}, \citenamefont {Mietschke}, \citenamefont {Zhang},
  \citenamefont {Yuan}, \citenamefont {Abel}, \citenamefont {H\"{u}hne},
  \citenamefont {Nielsch}, \citenamefont {Fompeyrine}, \citenamefont {Ding},\
  and\ \citenamefont {Schmidt}}]{ZhangY2016}%
  \BibitemOpen
  \bibfield  {author} {\bibinfo {author} {\bibfnamefont {Y.}~\bibnamefont
  {Zhang}}, \bibinfo {author} {\bibfnamefont {Y.}~\bibnamefont {Chen}},
  \bibinfo {author} {\bibfnamefont {M.}~\bibnamefont {Mietschke}}, \bibinfo
  {author} {\bibfnamefont {L.}~\bibnamefont {Zhang}}, \bibinfo {author}
  {\bibfnamefont {F.}~\bibnamefont {Yuan}}, \bibinfo {author} {\bibfnamefont
  {S.}~\bibnamefont {Abel}}, \bibinfo {author} {\bibfnamefont {R.}~\bibnamefont
  {H\"{u}hne}}, \bibinfo {author} {\bibfnamefont {K.}~\bibnamefont {Nielsch}},
  \bibinfo {author} {\bibfnamefont {J.}~\bibnamefont {Fompeyrine}}, \bibinfo
  {author} {\bibfnamefont {F.}~\bibnamefont {Ding}}, \ and\ \bibinfo {author}
  {\bibfnamefont {O.~G.}\ \bibnamefont {Schmidt}},\ }\href {\doibase
  10.1021/acs.nanolett.6b02523} {\bibfield  {journal} {\bibinfo  {journal}
  {Nano Letters}\ }\textbf {\bibinfo {volume} {16}},\ \bibinfo {pages} {5785}
  (\bibinfo {year} {2016})}\BibitemShut {NoStop}%
\bibitem [{\citenamefont {M\"{u}ller}\ \emph {et~al.}(2014)\citenamefont
  {M\"{u}ller}, \citenamefont {Bounouar}, \citenamefont {J\"{o}ns},
  \citenamefont {Gl\"{a}ssl},\ and\ \citenamefont {Michler}}]{MullerM2014}%
  \BibitemOpen
  \bibfield  {author} {\bibinfo {author} {\bibfnamefont {M.}~\bibnamefont
  {M\"{u}ller}}, \bibinfo {author} {\bibfnamefont {S.}~\bibnamefont
  {Bounouar}}, \bibinfo {author} {\bibfnamefont {K.~D.}\ \bibnamefont
  {J\"{o}ns}}, \bibinfo {author} {\bibfnamefont {M.}~\bibnamefont
  {Gl\"{a}ssl}}, \ and\ \bibinfo {author} {\bibfnamefont {P.}~\bibnamefont
  {Michler}},\ }\href {http://dx.doi.org/10.1038/nphoton.2013.377} {\bibfield
  {journal} {\bibinfo  {journal} {Nat. Photon.}\ }\textbf {\bibinfo {volume}
  {8}},\ \bibinfo {pages} {224} (\bibinfo {year} {2014})}\BibitemShut {NoStop}%
\bibitem [{\citenamefont {Bylander}\ \emph {et~al.}(2003)\citenamefont
  {Bylander}, \citenamefont {Robert-Philip},\ and\ \citenamefont
  {Abram}}]{Bylander2003}%
  \BibitemOpen
  \bibfield  {author} {\bibinfo {author} {\bibfnamefont {J.}~\bibnamefont
  {Bylander}}, \bibinfo {author} {\bibfnamefont {I.}~\bibnamefont
  {Robert-Philip}}, \ and\ \bibinfo {author} {\bibfnamefont {I.}~\bibnamefont
  {Abram}},\ }\href {\doibase 10.1140/epjd/e2002-00236-6} {\bibfield  {journal}
  {\bibinfo  {journal} {Eur. Phys. J. D}\ }\textbf {\bibinfo {volume} {22}},\
  \bibinfo {pages} {295} (\bibinfo {year} {2003})}\BibitemShut {NoStop}%
\bibitem [{\citenamefont {Widmann}\ \emph {et~al.}(2015)\citenamefont
  {Widmann}, \citenamefont {Portalupi}, \citenamefont {Lee}, \citenamefont
  {Michler},\ and\ \citenamefont {Gerhardt}}]{Widmann2015}%
  \BibitemOpen
  \bibfield  {author} {\bibinfo {author} {\bibfnamefont {M.}~\bibnamefont
  {Widmann}}, \bibinfo {author} {\bibfnamefont {S.}~\bibnamefont {Portalupi}},
  \bibinfo {author} {\bibfnamefont {S.-y.}\ \bibnamefont {Lee}}, \bibinfo
  {author} {\bibfnamefont {P.}~\bibnamefont {Michler}}, \ and\ \bibinfo
  {author} {\bibfnamefont {I.}~\bibnamefont {Gerhardt}},\ }\href
  {https://arxiv.org/abs/1505.01719v1} {\bibfield  {journal} {\bibinfo
  {journal} {arXiv preprint: 1505.01719}\ } (\bibinfo {year}
  {2015})}\BibitemShut {NoStop}%
\bibitem [{\citenamefont {Zieli\'{n}ska}\ \emph {et~al.}(2012)\citenamefont
  {Zieli\'{n}ska}, \citenamefont {Beduini}, \citenamefont {Godbout},\ and\
  \citenamefont {Mitchell}}]{Zielinska2011}%
  \BibitemOpen
  \bibfield  {author} {\bibinfo {author} {\bibfnamefont {J.~A.}\ \bibnamefont
  {Zieli\'{n}ska}}, \bibinfo {author} {\bibfnamefont {F.~A.}\ \bibnamefont
  {Beduini}}, \bibinfo {author} {\bibfnamefont {N.}~\bibnamefont {Godbout}}, \
  and\ \bibinfo {author} {\bibfnamefont {M.~W.}\ \bibnamefont {Mitchell}},\
  }\href {\doibase 10.1364/OL.37.000524} {\bibfield  {journal} {\bibinfo
  {journal} {Opt. Lett.}\ }\textbf {\bibinfo {volume} {37}},\ \bibinfo {pages}
  {524} (\bibinfo {year} {2012})}\BibitemShut {NoStop}%
\bibitem [{Note1()}]{Note1}%
  \BibitemOpen
  \bibinfo {note} {The software \protect \textit {ElecSus}~\cite {Zentile2015}
  is used to calibrate the conversion from coil current to magnetic
  field.}\BibitemShut {Stop}%
\bibitem [{\citenamefont {Kuhlmann}\ \emph {et~al.}(2013)\citenamefont
  {Kuhlmann}, \citenamefont {Houel}, \citenamefont {Ludwig}, \citenamefont
  {Greuter}, \citenamefont {Reuter}, \citenamefont {Wieck}, \citenamefont
  {Poggio},\ and\ \citenamefont {Warburton}}]{Kuhlmann2013}%
  \BibitemOpen
  \bibfield  {author} {\bibinfo {author} {\bibfnamefont {A.~V.}\ \bibnamefont
  {Kuhlmann}}, \bibinfo {author} {\bibfnamefont {J.}~\bibnamefont {Houel}},
  \bibinfo {author} {\bibfnamefont {A.}~\bibnamefont {Ludwig}}, \bibinfo
  {author} {\bibfnamefont {L.}~\bibnamefont {Greuter}}, \bibinfo {author}
  {\bibfnamefont {D.}~\bibnamefont {Reuter}}, \bibinfo {author} {\bibfnamefont
  {A.~D.}\ \bibnamefont {Wieck}}, \bibinfo {author} {\bibfnamefont
  {M.}~\bibnamefont {Poggio}}, \ and\ \bibinfo {author} {\bibfnamefont {R.~J.}\
  \bibnamefont {Warburton}},\ }\href {\doibase 10.1038/nphys2688} {\bibfield
  {journal} {\bibinfo  {journal} {Nat. Phys.}\ }\textbf {\bibinfo {volume}
  {9}},\ \bibinfo {pages} {570} (\bibinfo {year} {2013})}\BibitemShut {NoStop}%
\bibitem [{Note2()}]{Note2}%
  \BibitemOpen
  \bibinfo {note} {National Instruments NI PXI-7842R card}\BibitemShut
  {NoStop}%
\bibitem [{Note3()}]{Note3}%
  \BibitemOpen
  \bibinfo {note} {Calculated using $\protect \sqrt {\sigma _N^2-\protect
  \overline {N}}$ to exclude the detection shot noise. $\protect \overline {N}$
  is the average count number for $\geq $0.5~s binning times and $\sigma _N$ is
  the corresponding standard deviation.}\BibitemShut {Stop}%
\bibitem [{\citenamefont {Jung}\ and\ \citenamefont {Gweon}(2000)}]{Jung2000}%
  \BibitemOpen
  \bibfield  {author} {\bibinfo {author} {\bibfnamefont {H.}~\bibnamefont
  {Jung}}\ and\ \bibinfo {author} {\bibfnamefont {D.-G.}\ \bibnamefont
  {Gweon}},\ }\href {\doibase 10.1063/1.1150559} {\bibfield  {journal}
  {\bibinfo  {journal} {Rev. Sci. Instrum.}\ }\textbf {\bibinfo {volume}
  {71}},\ \bibinfo {pages} {1896} (\bibinfo {year} {2000})}\BibitemShut
  {NoStop}%
\bibitem [{\citenamefont {J\"ons}\ \emph {et~al.}(2017)\citenamefont {J\"ons},
  \citenamefont {Stensson}, \citenamefont {Reindl}, \citenamefont {Swillo},
  \citenamefont {Huo}, \citenamefont {Zwiller}, \citenamefont {Rastelli},
  \citenamefont {Trotta},\ and\ \citenamefont {Bj\"ork}}]{Jons2017}%
  \BibitemOpen
  \bibfield  {author} {\bibinfo {author} {\bibfnamefont {K.~D.}\ \bibnamefont
  {J\"ons}}, \bibinfo {author} {\bibfnamefont {K.}~\bibnamefont {Stensson}},
  \bibinfo {author} {\bibfnamefont {M.}~\bibnamefont {Reindl}}, \bibinfo
  {author} {\bibfnamefont {M.}~\bibnamefont {Swillo}}, \bibinfo {author}
  {\bibfnamefont {Y.}~\bibnamefont {Huo}}, \bibinfo {author} {\bibfnamefont
  {V.}~\bibnamefont {Zwiller}}, \bibinfo {author} {\bibfnamefont
  {A.}~\bibnamefont {Rastelli}}, \bibinfo {author} {\bibfnamefont
  {R.}~\bibnamefont {Trotta}}, \ and\ \bibinfo {author} {\bibfnamefont
  {G.}~\bibnamefont {Bj\"ork}},\ }\href {\doibase 10.1103/PhysRevB.96.075430}
  {\bibfield  {journal} {\bibinfo  {journal} {Phys. Rev. B}\ }\textbf {\bibinfo
  {volume} {96}},\ \bibinfo {pages} {075430} (\bibinfo {year}
  {2017})}\BibitemShut {NoStop}%
\bibitem [{\citenamefont {Zentile}\ \emph {et~al.}(2015)\citenamefont
  {Zentile}, \citenamefont {Keaveney}, \citenamefont {Weller}, \citenamefont
  {Whiting}, \citenamefont {Adams},\ and\ \citenamefont
  {Hughes}}]{Zentile2015}%
  \BibitemOpen
  \bibfield  {author} {\bibinfo {author} {\bibfnamefont {M.~A.}\ \bibnamefont
  {Zentile}}, \bibinfo {author} {\bibfnamefont {J.}~\bibnamefont {Keaveney}},
  \bibinfo {author} {\bibfnamefont {L.}~\bibnamefont {Weller}}, \bibinfo
  {author} {\bibfnamefont {D.~J.}\ \bibnamefont {Whiting}}, \bibinfo {author}
  {\bibfnamefont {C.~S.}\ \bibnamefont {Adams}}, \ and\ \bibinfo {author}
  {\bibfnamefont {I.~G.}\ \bibnamefont {Hughes}},\ }\href {\doibase
  10.1016/j.cpc.2014.11.023} {\bibfield  {journal} {\bibinfo  {journal}
  {Comput. Phys. Commun.}\ }\textbf {\bibinfo {volume} {189}},\ \bibinfo
  {pages} {162 } (\bibinfo {year} {2015})}\BibitemShut {NoStop}%
\end{thebibliography}
\end{document}